\documentclass[preprint2]{aastex631}

\shorttitle{A persistent disk wind and variable jet outflow in GX 13+1}
\shortauthors{Rogantini, Homan, Plotkin et al. (2025)}
\graphicspath{{./}{figures/}}

\DeclareMathAlphabet{\mathsc}{OT1}{cmr}{m}{sc}
\def\testbx{bx}%
\DeclareRobustCommand{\ion}[2]{%
\relax\ifmmode
\ifx\testbx\f@series
{\mathbf{#1\,\mathsc{#2}}}\else
{\mathrm{#1\,\mathsc{#2}}}\fi
\else\textup{#1\,{\mdseries\textsc{#2}}}%
\fi}

\newcommand{\gx}{GX~13+1\xspace}
\newcommand{\asca}{{\it ASCA}\xspace}
\newcommand{\chandra}{{\it Chandra}\xspace}

\newcommand{\exosat}{{EXOSAT}\xspace}

\newcommand{\nicer}{{\it NICER}\xspace}

\newcommand{\xte}{{\it RXTE}\xspace}

\newcommand{\vla}{{\it VLA}\xspace}

\newcommand{\xmm}{XMM{\it-Newton}\xspace}
\newcommand{\xrism}{{\it XRISM}\xspace}

\newcommand{\vel}{km\,s$^{-1}$\xspace}

\newcommand{\kms}{\ensuremath{\mathrm{km\ s^{-1}}}\xspace}

\newcommand{\ledd}{$L_{\rm Edd}$}

\newcommand{\NH}{\ensuremath{N_{\mathrm{H}}}\xspace}

\newcommand{\spex}{{\textsc{Spex}}\xspace}
\newcommand{\pion}{{\texttt{pion}}\xspace}

\newcommand{\logxi}{\ensuremath{\log \xi}\xspace}
\newcommand{\logxierg}{\ensuremath{\log (\xi/\rm erg\: cm\: s^{-1})}\xspace}
\newcommand{\vout}{\ensuremath{v_{\rm out}}\xspace}
\newcommand{\vturb}{\ensuremath{v_{\rm turb}}\xspace}

\newcommand{\fexxv}{\ion{Fe}{xxv}\xspace}
\newcommand{\fexxvi}{\ion{Fe}{xxvi}\xspace}

\usepackage{tabularx}
\usepackage{graphicx}
\usepackage[referable]{threeparttablex}
\usepackage{txfonts}
\usepackage{multirow}
\usepackage{amssymb}
\usepackage{units}
\usepackage{float}
\usepackage[caption = false]{subfig}
\usepackage{natbib}
\bibpunct{(}{)}{;}{a}{}{,}
\usepackage{url}
\usepackage{array}
\usepackage{xspace}
\usepackage{enumitem,booktabs}

\definecolor{orchid}{RGB}{218,30,211}

\makeatletter

\begin{document}

\title{A persistent disk wind and variable jet outflow in the neutron-star low-mass X-ray binary GX 13+1}

\author[0000-0002-0786-7307]{Daniele Rogantini}

\affiliation{Department of Astronomy and Astrophysics, University of Chicago, 5640 S Ellis Ave, Chicago, IL 60637, USA}

\author[0000-0001-8371-2713]{Jeroen Homan}
\affiliation{Eureka Scientific, Inc., 2452 Delmer Street, Oakland, CA 94602, USA}

\author[0000-0002-7092-0326]{Richard M. Plotkin}
\affiliation{Department of Physics, University of Nevada, Reno, NV 89557, USA}
\affiliation{Nevada Center for Astrophysics, University of Nevada, Las Vegas, NV 89154, USA}

\author[0000-0003-0746-795X]{Maureen van den Berg}
\affiliation{Center for Astrophysics, Harvard \& Smithsonian, 60 Garden Street, Cambridge, MA 02138, USA}

\author[0000-0003-3124-2814]{James Miller-Jones}
\affiliation{International Centre for Radio Astronomy Research, Curtin University, GPO Box U1987, Perth WA 6845, Australia}

\author[0000-0002-8247-786X]{Joey Neilsen}
\affiliation{Department of Physics, Villanova University, 800 Lancaster Avenue, Villanova, PA 19085, USA}

\author[0000-0001-8804-8946]{Deepto Chakrabarty}
\affiliation{Department of Physics, Massachusetts Institute of Technology, Cambridge, MA 02139, USA}
\affiliation{MIT Kavli Institute for Astrophysics and Space Research
Massachusetts Insitute of Technology, Cambridge, MA 02139, USA}

\author{Rob P. Fender}
\affiliation{Astrophysics, Department of Physics, University of Oxford, Keble Road, Oxford OX1 3RH, UK}

\author{Norbert Schulz}
\affiliation{MIT Kavli Institute for Astrophysics and Space Research
Massachusetts Insitute of Technology, Cambridge, MA 02139, USA}

\begin{abstract}

In low-mass X-ray binaries (LMXBs), accretion flows are often associated with either jet outflows or disk winds. Studies of LMXBs with luminosities up to roughly $20\%$ of the Eddington limit indicate that these outflows generally do not co-occur, suggesting that disk winds might inhibit jets. However, previous observations of LMXBs accreting near or above the Eddington limit show that jets and winds can potentially coexist. To investigate this phenomenon, we carried out a comprehensive multi-wavelength campaign (using \vla, \chandra/HETG, and \nicer) on the near-Eddington neutron-star Z source LMXB \gx. \nicer and \chandra/HETG observations tracked \gx across the entire Z-track during high Eddington rates, detecting substantial resonance absorption features originating from the accretion disk wind in all X-ray spectra, which implies a persistent wind presence. Simultaneous \vla observations captured a variable radio jet, with radio emission notably strong during all flaring branch observations—contrary to typical behavior in Z-sources—and weaker when the source was on the normal branch. Interestingly, no clear correlation was found between the radio emission and the wind features. Analysis of \vla radio light curves and simultaneous \chandra/HETG spectra demonstrates that an ionized disk wind and jet outflow can indeed coexist in \gx, suggesting that their launching mechanisms are not necessarily linked in this system. 
\end{abstract}

\renewcommand{\thefootnote}{}%
\footnotetext{Accepted for publication in The Astrophysical Journal. \\ © 2025. All rights reserved.}%
\renewcommand{\thefootnote}{\arabic{footnote}}

\section{Introduction}  \label{sec:intro}

In low-mass X-ray binaries (LMXBs), accretion processes are often associated with either highly collimated jet outflows or more equatorial disk winds. Jets are typically relativistic and launched near the compact object \citep{Fender06, Fender10}, while disk winds are slower and originate farther out \citep{Neilsen13, DiazTrigo16}. After hints in the black hole (BH) LMXB H1743--332 that jets and disk winds are causally related \citep{Miller06b, Miller08b}, \citet{Neilsen09} found that in the BH LMXB GRS 1915+105 the jet outflows are typically quenched when strong disk winds are present. \citet{Neilsen09} interpreted this anti-correlation between jets and disk winds as the latter being able to carry away sufficient mass to suppress the flow of matter into the jet.

A large study by \citet{Ponti12a} on transient BH LMXBs revealed that disk winds in high-inclination systems (where they are most easily detected) are mainly found in spectrally soft X-ray states. Combined with the fact that radio jets in BH LMXBs are predominantly found in the spectrally hard X-ray state \citep{Fender09}, this further suggested that jets and disk winds are to a large extent mutually exclusive. Using Fe K absorption lines, which are often used to trace disk winds (or disk atmospheres), \citet{Ponti14, Ponti15} found the same dependence on spectral state in two high-inclination neutron-star (NS) LMXBs as in BH systems. However, the spectral resolution was too low to constrain outflow speeds in these NS sources.

The above findings on the state dependence of disk winds in BH and NS LMXBs came mostly from systems with luminosities below approximately $20\%$ of the Eddington luminosity (\ledd). The only exception to the overall trend seen in the sample of \citet{Ponti12a} is GRS 1915+105, which has shown disk winds in a luminous hard state ($\gtrsim$0.3 \ledd); in that observation, GRS 1915+105 produced a jet and disk wind at the same time  \citep{Lee02, Neilsen09}. This suggests a possible shift in the jet-wind relationship as luminosities approach \ledd.

To further investigate this idea, \citet{Homan16} studied the jet-wind interaction in luminous ($>0.5$ \ledd) BH and NS LMXBs. Although they did not find another luminous LMXB with strictly simultaneous detections of radio jets and disk winds, they found strong circumstantial evidence that four other sources are able to launch disk winds and jets at the same time as well (e.g.\ observed in the same spectral state and/or very close in time). These sources were the BH LMXB V404 Cyg \citep[see also][]{King15b} and the NS LMXBs Sco X-1, Cir X-1, and \gx. The presence of jets in these sources was in all cases deduced from radio observations, while disk winds were inferred either from high-resolution X-ray spectra or near-infrared spectra. These findings strongly suggest that when LMXBs approach or exceed \ledd, jets and disk winds can coexist. This paints a different picture of the relation between jets and disk winds than the one that has emerged from lower-luminosity sources.

Of the five sources for which simultaneous jets and disk winds have been observed or implied, \gx\ is the only one that has consistently shown wind signatures in all of its \chandra\ observations (seven in total). \gx\ is a persistently bright NS LMXB that was initially classified as a so-called atoll source (i.e., a low-luminosity NS LMXB) by \citet{Hasinger89} based on \exosat\ data. However, an analysis by \citet{Fridriksson15} showed that, at least during the lifetime of \xte\ (1995--2012), \gx\ exhibited properties similar to Z sources, a subclass of NS-LMXBs with luminosities near or above \ledd. Recently, \nicer\ observed \gx\ for 22 days between February and April 2024, covering the entire Z-track \citep{Kaddouh24}. X-ray spectral-polarimetric observations suggest a misalignment between the neutron star spin axis and the orbital plane \citep{Bobrikova24a,Bobrikova24b}. Periodic (24.27 d) absorption dips have been observed in the X-ray light curves of \gx\ \citep{Iaria14, dAi14, DiMarco25}, implying a high inclination (65$^\circ$--70$^\circ$). As in other high-inclination systems, evidence for an ionized outflow was found in \gx, with {\it ASCA} and \xmm\ \citep{Ueda01, Sidoli02}. Later observations with \chandra\ revealed that the absorption lines from this ionized outflow were significantly blue-shifted, indicating the presence of a disk wind with outflow speeds up to  $\sim$800 \vel\  \citep{Ueda04, Madej14}. \citet{Ueda04} suggest that the wind is driven by radiation pressure and they infer a mass outflow rate of at least 0.7$\times10^{18}$ g\,s$^{-1}$, which is comparable to the mass accretion rate. This illustrates that winds are important to the overall dynamics of the accretion flow in \gx. 

\gx\  is highly variable in the radio \citep{Garcia88, Homan04} with luminosities similar to those of the other Z sources \citep{Fender00}, indicating the presence of strong jet outflows. Like in most of the other Z sources \citep[see, e.g.,][GX~17+2, Cyg~X-2 and Sco~X-1]{Penninx88,Hjellming90a,Hjellming90b,Migliari07} the evolution of the radio flux of \gx\ suggests that the jet is strongest on the so-called normal- (NB) and horizontal branches (HB) of its Z track and strongly suppressed on its flaring branch (FB) \citep[][]{Homan04}. 

\cite{Homan16} studied the X-ray spectral state dependence of the disk wind in \gx. To investigate this dependence they used \xte observations that were performed simultaneously with six of the seven \chandra observations. All of the archival \chandra observations of \gx were made when the source was on the horizontal- and normal- branches. This also happens to be part of the Z track on which the jets in \gx are the strongest. Although the \chandra and radio observations were not made simultaneously, they suggest that the horizontal/normal branches in \gx can give rise to disk winds as well as radio jets. Notably, the radio flux density is much weaker on the flaring branch, and it is not clear if the radio emission on that branch is from a (weak but active) compact jet or from previously ejected material. Unfortunately, the \chandra data did not yield insights into the presence of disk winds on the flaring branch. Signatures of an ionized outflow were also seen in (all) lower-resolution \asca and \xmm\ spectra of \gx. The presence of flaring in the light curves of some of the \xmm\ observations \citep{DiazTrigo12}, suggests that the ionized outflow is also present on the flaring branch, albeit with poorly constrained outflow properties. We also note that in addition to narrow absorption lines, the \xmm\ spectra revealed a broad Fe emission line. 

We present results from a multi-wavelength campaign (Joint \chandra proposal 19400246) designed to explore the relationship between the disk wind and radio jet in the bright NS X-ray binary, GX 13+1. Section \ref{sec:log} outlines the \chandra, \nicer, and Karl G. Jansky Very Large Array (\vla) observations and data reduction. Section \ref{sec:results} describes the hardness-intensity analysis, broadband spectral fitting, wind photoionization modeling, and radio jet analysis. We discuss and interpret our findings in Section \ref{sec:discussion} and summarize our conclusions in Section \ref{sec:conclusion}.

\section{Observations and Data Reduction} \label{sec:log}
\gx was observed by \chandra, \nicer, and \vla between May 2018 and May 2019. The observation log is presented in Table \ref{tab:log}. Four epochs (1, 2, 4, and 5) included simultaneous \chandra, \nicer, and \vla observations, while epoch 3 consisted of a brief \nicer observation only. Light curves for all observations in this work are shown in Figure \ref{fig:lc}. A description of the observations and data reduction procedures follows.

%
\begin{table*}
\begin{center}
\caption{\chandra/HETG, \nicer and \vla observation log.}   
\begin{tabular}{ c c c c c c c } 
\hline
\noalign{\vskip 0.5mm}
Epoch & Instrument & ObsID & Date & Start Time & Exp. Time & Count Rate  \\
 &  &  & yyyy-mm-dd & UTC & Integrated & cts/s  \\ 
\noalign{\vskip 0.5mm}
\hline
\noalign{\vskip 0.5mm}
\multirow{3}{*}{1}  & \chandra/HETG & 20191          & 2018-05-10 & 06:21:06 & $24.3\:\rm ks$   &  102.44  \\ 
                    & \nicer        & 1108020103     & 2018-05-10 & 07:46:28 & $3.6\:\rm ks$    &  906.12 \\ 
                    & \vla          & SJ0246.sb35351382    & 2018-05-10 & 07:30:00 &  5 hrs           &  --  \\
\noalign{\vskip 0.5mm}
\hline 
\noalign{\vskip 0.5mm}
\multirow{3}{*}{2}  & \chandra/HETG & 20192         & 2018-05-26 & 07:09:48 & $22.3\:\rm ks$   &  88.03 \\
                    & \nicer        & 1108020104    & 2018-05-26 & 08:38:13 & $4.6\:\rm ks$    & 724.12  \\
                    & \vla          & SJ0246.sb35426119    & 2018-05-26 & 07:15:00&   5 hrs           & --  \\
\noalign{\vskip 0.5mm}
\hline
\noalign{\vskip 0.5mm}
3                   & \nicer        & 1108020105    & 2018-07-23 & 03:46:20 & $0.9\:\rm ks$    &  767.42 \\
\noalign{\vskip 0.5mm}
\hline
\noalign{\vskip 0.5mm}
\multirow{3}{*}{4}  & \chandra/HETG & 20193         & 2019-02-17 & 11:54:29 & $24.3\:\rm ks$   &  116.54 \\
                    & \nicer        & 1108020106    & 2019-02-17 & 12:11:54 & $1.0\:\rm ks$    &  1000.08 \\
                    & \vla          & SJ0246.sb36280238    & 2019-02-17 & 13:09:55 &  5 hrs           &  -- \\
\noalign{\vskip 0.5mm}
\hline
\noalign{\vskip 0.5mm}
\multirow{3}{*}{5}  & \chandra/HETG & 20194         & 2019-05-30 & 03:44:31 & $24.6\:\rm ks$   & 99.82  \\ 
                    & \nicer        & 2108020101    & 2019-05-30 & 05:10:14 & $2.4\:\rm ks$    & 818.77 \\
                    & \vla          & SJ0246.sb36689258    & 2019-05-30 & 06:00:00 & 5 hrs            & --  \\
\noalign{\vskip 0.5mm}
\hline
\label{tab:log}
\end{tabular}
\end{center}
\end{table*}

\subsection{Chandra} \label{sec:data_reduction_chandra}
\chandra \citep{Weisskopf02} observed \gx four times between May 2018 and May 2019 with an average exposure time of approximately 24 ks per observation. All observations were performed using the High Energy Transmission Grating Spectrometer (HETGS) in Timed Exposure mode \citep{Canizares05}. We extracted the High Energy Grating (HEG) and Medium Energy Grating (MEG) spectra using the \texttt{CIAO} X-ray data analysis package \citep[version 4.16;][]{Fruscione06} and the latest calibration (CALDB, version 4.11), following the \chandra Gratings Catalog and Archive (TGCat) processing procedures \citep{Huenemoerder11}. To mitigate CCD pileup effects, which significantly impact bright sources like \gx, we employed a 350-row subarray centered on the source and activated only four CCDs. This configuration effectively reduced the detector frame time, allowing us to obtain pileup-free first-order HEG spectra \citep{Schulz16}. For the MEG spectra, we excluded the portion above 2 keV, where pileup effects occur. The positive and negative first-order spectra for each instrument were combined using the \textsc{combine\_grating\_spectra} tool in CIAO. HEG and MEG spectra were fitted simultaneously using a cross-calibration constant, which remained close to 1, indicating less than 5\% calibration variation between the instruments. HEG and MEG data were analyzed over the $1.2-10$\,keV and $1-2$\,keV energy ranges ($1.23-10.3$\,\AA\ and $6.2-12.4$\,\AA), respectively.

\subsection{NICER} \label{sec:nicer}
Each \chandra observation was accompanied by short \nicer observations (see Table \ref{tab:log}) aimed at constructing hardness-intensity diagrams to track the source’s spectral branch position and improve constraints on continuum parameters. A further \nicer observation from July 2018 was included to extend coverage of the Z track.

All \nicer data were reprocessed using the {\tt nicerl2} tool in HEASOFT 6.33 with the latest calibration files (20240206.tar) and default filters. Light curves were extracted from active Focal Plane Modules across various energy bands with a time resolution of 16\,s. Background corrections were unnecessary, as background rates were negligible in the selected energy bands. To track the source across spectral branches, we created a hardness-intensity diagram (HID), using count rates in the 0.5 -- 10 keV band for intensity and the 4 -- 10 keV/0.5 -- 2 keV band count-rate ratio for hard color. 

A single, time-averaged spectrum was extracted per \nicer observation using the HEASOFT tool {\tt nibackgen3C50}, which also generated background spectra \citep{remillard2022}. Response and ancillary files were created using {\tt nicerrmf} and {\tt nicerarf}, respectively.

\subsection{Very Large Array} \label{sec:vla}
The \vla observed simultaneously with \chandra over four epochs, through program SJ0246.  The \vla was in its most extended A configuration during epochs 1 and 2,  
was being moved from C to B configuration during epoch 4, 
and was in B configuration during epoch 5.    
Observations were taken in subarray mode, with approximately half of the antennas observing in C-band (with 2$\times$1 GHz basebands centered at 5.25 and 7.45 GHz) and the other half in Ku-band (with 2$\times$1 GHz basebands centered at 14.4 and 17.2 GHz), thereby providing simultaneous spectral information from  5.25 through 17.2 GHz.  Each epoch lasted for five hours, with a time on source of $\approx$4 hours at 5.25/7.45 GHz and $\approx$3 hours at 14.4/17.2 GHz (before flagging).  In all cases, 3C 286 was used as the bandpass calibrator and for setting the flux density scale.  The phase calibrators J1820-2528 and J1825-1718 were used to determine the complex gain solutions at 5.25/7.45 and 14.4/17.2 GHz, respectively, and a reference pointing scan was performed every $\approx$60 min for the 14.4/17.2 GHz subarray. 

Data were reduced using the Common Astronomy Software Applications ({\tt CASA}) v6.1.2 \citep{Casa22}. Calibration was performed using the standard \vla calibration pipeline v5.4. Additional minor flagging was applied after manual inspection. Our ultimate goal was to create light curves on sub-hour timescales. We generated preliminary images for the first 10 scans of each epoch using {\tt tclean} to create `reconnaissance' images at each of the four central frequencies (5.25, 7.45, 14.4 and 17.2 GHz) using two Taylor terms ({\tt nterms=2}) to account for the wide fractional bandwidth. We adopted Briggs weighting with {\tt robust=1} (5.25/7.45 GHz) and {\tt robust=0} (14.4/17.2 GHz), as a compromise between maximizing sensitivity and minimizing sidelobes.

\gx was detected at all four frequencies in each epoch, and we fit a two-dimensional Gaussian to the target in each image (with the task {\tt imfit}) to measure a preliminary flux density (which ranged from $\approx$0.2--7.0 mJy bm$^{-1}$, depending on the epoch and frequency). {\tt imfit} also reported that radio emission from \gx was unresolved on all four epochs. From these preliminary flux densities and the root-mean-square noise ($\sigma_{\rm rms}$) measured from a source-free region of each image, we determined the optimal time bin for our final light curves to allow $>$5$\sigma_{\rm rms}$ detections per time bin at each frequency (although, we note that $\gtrsim$20-30$\sigma_{\rm rms}$ detections were more common). Final light curves were created with 10, 30, 10, and 5-minute bins for epochs 1, 2, 4, and 5, respectively. Peak flux densities of \gx were then measured in each image using {\tt imfit}. Typical $\sigma_{\rm rms}$ values within each time bin were $\approx$0.03 (0.05), 0.02 (0.04), 0.04 (0.05), 0.05 (0.07) mJy at 5.25/7.45 (14.4/17.2) GHz for epochs 1, 2, 4, and 5, respectively.  

Finally, over each time bin, radio spectral indices, $\alpha$, defined as $f_\nu \propto \nu^\alpha$ (where $f_\nu$ is flux density at frequency $\nu$), were derived from the four flux densities using a linear fit in log space. Uncertainties on the best-fit values of $\alpha$ were determined through Monte Carlo simulations, where we perturbed each measured flux density by randomly drawing a new flux density from a normal distribution with a mean and standard deviation equal to the measured flux density and uncertainty at each frequency; we also perturbed each central frequency by randomly drawing a new frequency over a uniform distribution covering the 1 GHz bandwidth.  We then performed a linear regression on each simulated spectrum, and we repeated 1000 times. Uncertainties on $\alpha$ were calculated as the 68\% confidence level of the resulting distribution of 1000 simulated $\alpha$ values.

\begin{figure*}[t]
\centering
\includegraphics[width=\textwidth]{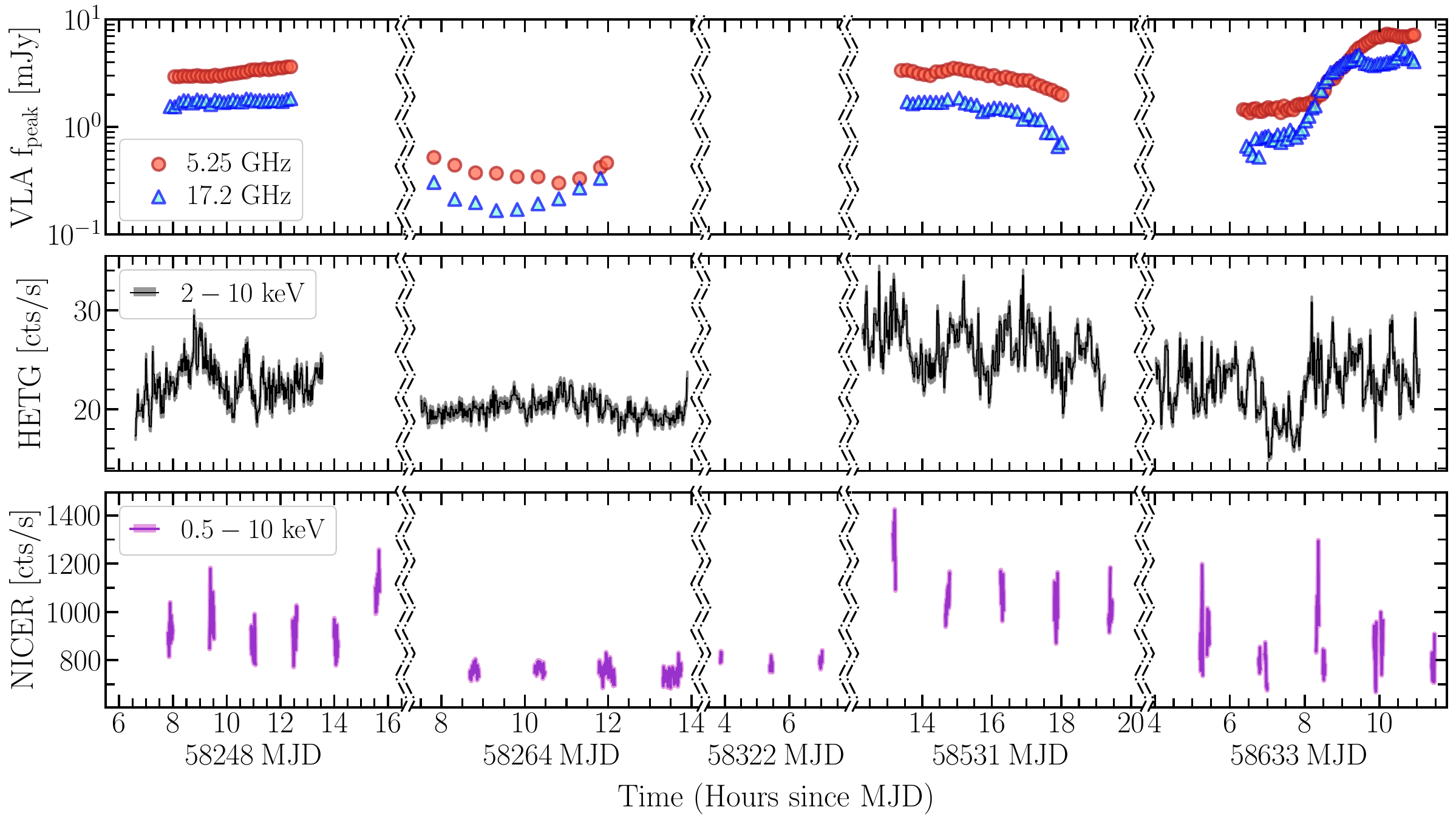}\\

\caption{Multi wavelength light curves. 
{\it Top}: \vla light curves at two different frequencies, 5.25 GHz in red circles and 17.2 GHz in blue triangles. {\it Middle}: \chandra/HETG light curve in the 2-10 keV energy range with a time bin of 100 seconds. 
{\it Bottom:} \nicer light curve in the 0.5-10 keV energy range with a time bin of 60 seconds.}
\label{fig:lc}
\end{figure*}

\section{Results} \label{sec:results}
We analysed the HETG and \nicer spectra with the spectral fitting program SPEX v3.07.03 \citep{Kaastra22} adopting Cash statistics \citep{Cash79}. Spectra were converted from \textsc{ogip} to SPEX format using \textsc{trafo} and data were binned using optimized binning, {\tt obin}, in SPEX \citep{Kaastra16}. The best values of our fits are presented with 1$\sigma$ uncertainties.  We report here the results of our X-ray/radio variability and spectral analyses.

\subsection{X-ray Light Curves and HID} 
The middle and bottom panels of Figure \ref{fig:lc} show the 2--10 keV \chandra/HETG and 0.5--10 keV \nicer light curves of \gx, respectively. \gx has exhibited type-I X-ray bursts \citep{matsuba1995} and absorption dips \citep{Iaria14,DiMarco25} previously, but none were seen in our five-epoch light curves. In epochs 1, 4, and 5, \gx showed notable X-ray variability, while the count rates were relatively stable during epochs 2 and 3.

\begin{figure}
\centerline{\includegraphics[width=8cm]{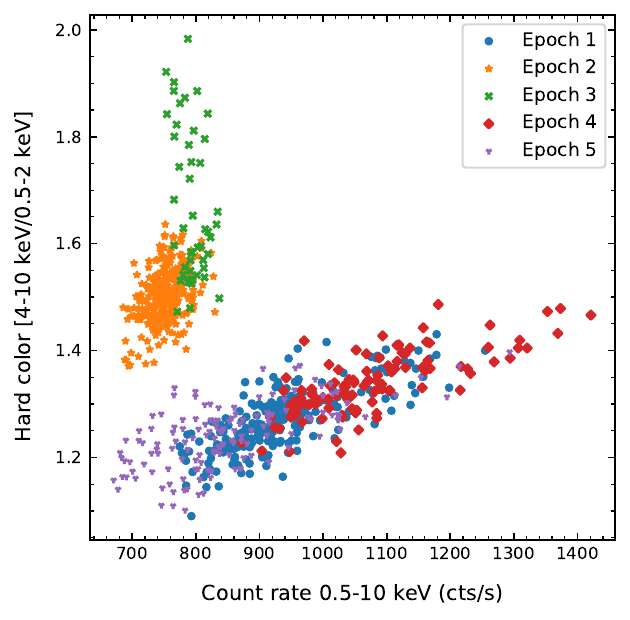}}
\caption{Hardness-intensity diagram of the five \nicer observations of \gx. For intensity, we used the count-rate in the 0.5--10 keV band, and for the hard color the ratio of the count rates in the 4--10 keV and 0.5--2 keV bands. Each data point represents a 16\,s time interval.}\label{fig:hid}
\end{figure}

Figure \ref{fig:hid} displays the \nicer HID, with each epoch differentiated by color and marker type. The HID tracks of \gx are known to drift and change in shape on time scales as short as a few days \citep{Fridriksson15}. Although our observations span over a year, the individual segments result in a track with only minor shifts, although some gaps are present. The overall shape of the track resembles the $\nu$-shaped Z tracks of Sco X-1 and GX 17+2, with vertically oriented normal and horizontal branches. Epochs 1, 4, and 5 mostly trace out the flaring branch, epoch 2 the upper normal branch, and epoch 3 the horizontal branch. This indicates that we managed to observe \gx on the flaring branch with \chandra for the first time.

\subsection{Broad-band X-ray spectra}

%
\begin{table*}
\caption{The spectral analysis results of the HETG spectra of \gx: best-fit values for the continuum and two ionized outflow components (W1 and W2) parameters.
}
\label{tab:results}   
\centering
\begin{threeparttable}        
\begin{tabular}{c c c | c c c c }     
\hline\hline       
\noalign{\vskip 0.75mm}
Comp. & Par. & Units & Epoch 1 & Epoch 2 & Epoch 4 & Epoch 5 \\
\noalign{\vskip 0.75mm}
\hline
\noalign{\vskip 0.75mm}              
\multicolumn{7}{c}{Continuum} \\
\noalign{\vskip 0.75mm}
\hline
\noalign{\vskip 0.25mm}  
\texttt{hot}$^\dagger$ & $N_{\rm H}$ & $10^{22}\:\rm cm^{-2}$ & $3.36\pm0.03$ & 
$2.93\pm0.03$ &
$2.74\pm0.03$ & 
$2.94\pm0.03$ \\
\noalign{\vskip 0.25mm}
\hline
\noalign{\vskip 0.25mm} 
\multirow{2}{*}{\texttt{bb}} & Norm & $ 10^{12}\:\rm cm^2 $ & $1.3_{-0.6}^{+1.8}$ & 
$0.6_{-0.4}^{+1.1}$ & 
$2.2_{-1.1}^{+2.2}$ &
$0.8_{-0.6}^{+1.3}$ \\
& $k$T & keV & $2.2\pm0.2$ &
$2.1_{-0.2}^{+0.8}$ & 
$2.1_{-0.1}^{+0.2}$ & 
$1.9_{-0.2}^{+0.5}$ \\
\noalign{\vskip 0.25mm}
\hline
\noalign{\vskip 0.25mm}  
\multirow{2}{*}{\texttt{mbb}} & Norm & $ 10^{26}\:\rm cm^{1/2}$ & $1.1_{-0.2}^{+1.0}$ &
$0.7_{-0.2}^{+0.8}$ & 
$1.8_{-0.7}^{+1.5}$ & 
$1.4\pm0.2$ \\
& $k$T & keV & $1.66\pm0.07$ & 
$2.05_{-0.18}^{+0.11}$ & 
$1.65\pm0.06$  & 
$1.72_{-0.08}^{+0.06}$ \\
\noalign{\vskip 0.25mm}
\hline
\noalign{\vskip 0.25mm}  
\texttt{amol} & $N_{\rm dust}$ & $10^{18}\:\rm cm^{-2}$ & $1.23\pm0.04$ & 
$1.47\pm0.04$ & 
$1.45\pm0.04$ &
$1.16\pm0.04$ \\
\noalign{\vskip 0.25mm}
\hline
\noalign{\vskip 0.25mm}  
\multirow{3}{*}{\texttt{gaus}} & Norm & $10^{44}\:\rm ph/s$ & $0.28\pm0.04$ & 
$0.30\pm0.03$ & 
$0.30\pm0.04$ & 
$0.21\pm0.03$ \\
& $E_{0}$ & keV & $6.54\pm0.06$ &
$6.59\pm0.04$ & 
$6.44\pm0.09$ &
$6.58\pm0.04$ \\
& FWHM & keV & $1.02_{-0.20}^{+0.24}$ & 
$0.80\pm0.20$ & 
$1.10\pm0.25$ & 
$0.59\pm0.12$ \\
\noalign{\vskip 0.25mm}
\hline
\noalign{\vskip 0.75mm}
\multicolumn{7}{c}{ionized disk wind} \\
\noalign{\vskip 0.75mm}
\hline
\noalign{\vskip 0.25mm}
\multirow{4}{*}{\texttt{pionW1}} & $N_{\rm H}$ & $10^{23}\:\rm cm^{-2}$ & $1.8_{-0.6}^{+1.0}$ &
$2.0_{-1.1}^{+1.5}$ & 
$0.6_{-0.2}^{+0.5}$ &
$5.3\pm1.2$ \\
& \multicolumn{2}{c|}{\logxierg}  & $4.0\pm0.1$ & 
$3.8_{-0.2}^{+0.1}$ & 
$3.8\pm0.2$ &
$3.83\pm0.05$ \\
& \vturb & \kms & $130_{-30}^{+70}$ &
$40\pm20$  & 
$110\pm40$ &
$40\pm10$ \\
& \vout & \kms & $430_{-40}^{+120}$ &
$280\pm30$ & 
$500\pm50$ &
$440\pm50$ \\
\noalign{\vskip 0.25mm}
\hline
\noalign{\vskip 0.25mm}  
\multirow{4}{*}{\texttt{pionW2}} & $N_{\rm H}$ & $10^{23}\:\rm cm^{-2}$ & $4.8_{-2.7}^{+9.1}$ &
$5.8_{-1.6}^{+1.2}$ & 
$9.9_{-3.9}^{+2.5}$ &
$3.7_{-1.0}^{+1.4}$ \\
& \multicolumn{2}{c|}{\logxierg}  & $4.7_{-0.2}^{+0.3}$ &
$4.5_{-0.2}^{+0.7}$ & 
$4.7\pm0.1$ &
$4.3\pm0.1$ \\
& \vturb & \kms & $130\pm30$ &
$110_{-20}^{+350}$ & 
$200\pm30$  &
$180\pm30$ \\
& \vout & \kms & $1220_{-80}^{+170}$ & 
$400\pm100$ &
$1100_{-70}^{+70}$ &
$990_{-130}^{+50}$ \\
\noalign{\vskip 0.25mm}
\hline
\noalign{\vskip 0.25mm}
\multicolumn{2}{c}{$F_{\rm observed;\: 2\mbox{-}10\:\rm keV}$} & $\rm 10^{-9}\: erg\:s^{-1}\:cm^{-2}$ & $6.2\pm0.9$& $5.5\pm0.8$ & $7.1\pm0.9$& $5.8\pm0.6$ \\
\multicolumn{2}{c}{$F_{\rm intrinsic;\: 2\mbox{-}10\:\rm keV}$} & $\rm 10^{-8}\: erg\:s^{-1}\:cm^{-2}$ & $1.4\pm0.3$& $1.4\pm0.4$ & $2.5\pm0.6$& $1.7\pm0.4$ \\
\noalign{\vskip 0.25mm}
\hline
\noalign{\vskip 0.25mm}
\multicolumn{3}{c|}{$C$stat/d.o.f.} & 2289/2157 & 2207/2109 & 2548/2159 & 2447/2152\\
\noalign{\vskip 0.25mm}
\hline
\end{tabular}
\begin{tablenotes}
\item $^\dagger$ The temperature of the cold neutral absorption is fixed to 0.008~eV ($\sim 100 K$) to mimic the cold, neutral ISM.
\end{tablenotes}
\end{threeparttable}
\end{table*}

As our main goal for the hereby presented programs is to study the evolution of the absorption line from the disk wind, it is necessary to accurately model the broadband continuum. This provides a proper spectral energy distribution (SED) necessary for the photoionisation modelling and helps prevent absorber models from fitting residuals due to incorrect continuum modeling rather than true disk wind features. We employed phenomenological models for the various emission components and applied a physical model, \pion \citep{Mehdipour16}, for ionized absorption from the wind. This model calculates the ionization balance dynamically within \spex using the current continuum model as the SED.

We began by fitting the \nicer and HETG spectra simultaneously, focusing on accurately capturing the continuum shape of the source. We tested several combinations of continuum components available in \spex (disc blackbody, power law, blackbody, modified blackbody) corrected for neutral Galactic absorption using the \texttt{hot} component which mimics a cold plasma in collisional ionization equilibrium \citep{dePlaa04}. While a few combinations showed similar statistical significance, we selected the phenomenological model consisting of a blackbody, \texttt{bb}, and a modified blackbody, \texttt{mbb}\footnote{This model describes the spectrum of a black body modified by coherent Compton scattering, which provides a better fit for certain astrophysical cases, such as accretion disk spectra. The formulae used are given by \citet{Kaastra89}.}, which yielded the lowest C-statistic value. This model, corrected for foreground neutral absorption, describes all five HETG/\nicer epochs well. After determining the best-fit continuum model, we restricted the spectral analysis to the HETG data to exploit the higher spectral resolution of the gratings. In addition, since \nicer overlaps with only a limited portion of the \chandra exposure, spectral differences between the two instruments may arise due to intra-observation variability.

Pronounced residuals near the Mg and Si K-edges (at $E=1.30$~keV and 1.84~keV, respectively) suggest the presence of interstellar dust along the line of sight toward \gx. Including a dust model is essential to accurately determine the continuum shape and avoid confusion between dust and wind absorption features. Thus, we corrected the continuum for cosmic dust extinction using the \texttt{amol} component \citep{Pinto10}. The recent update to this model incorporates a range of dust extinction models for both the Mg and Si K-edges \citep{Rogantini19,Zeegers19}. While a comprehensive analysis of dust properties is beyond the scope of this paper, we adopted amorphous olivine, $\rm MgFeSiO_{4}$, as a representative choice. \cite{Rogantini20} demonstrated that this silicate compound is best suited for large column densities, such as those observed toward \gx, and accurately captures the fine features near the Mg and Si K-edges. Adding the \texttt{amol} component to the fit resulted in a significant improvement to the C-statistic, with $\Delta C\rm stat > 140$ for each HETG observation.

\begin{figure*}[ht]
\includegraphics[width=\textwidth]{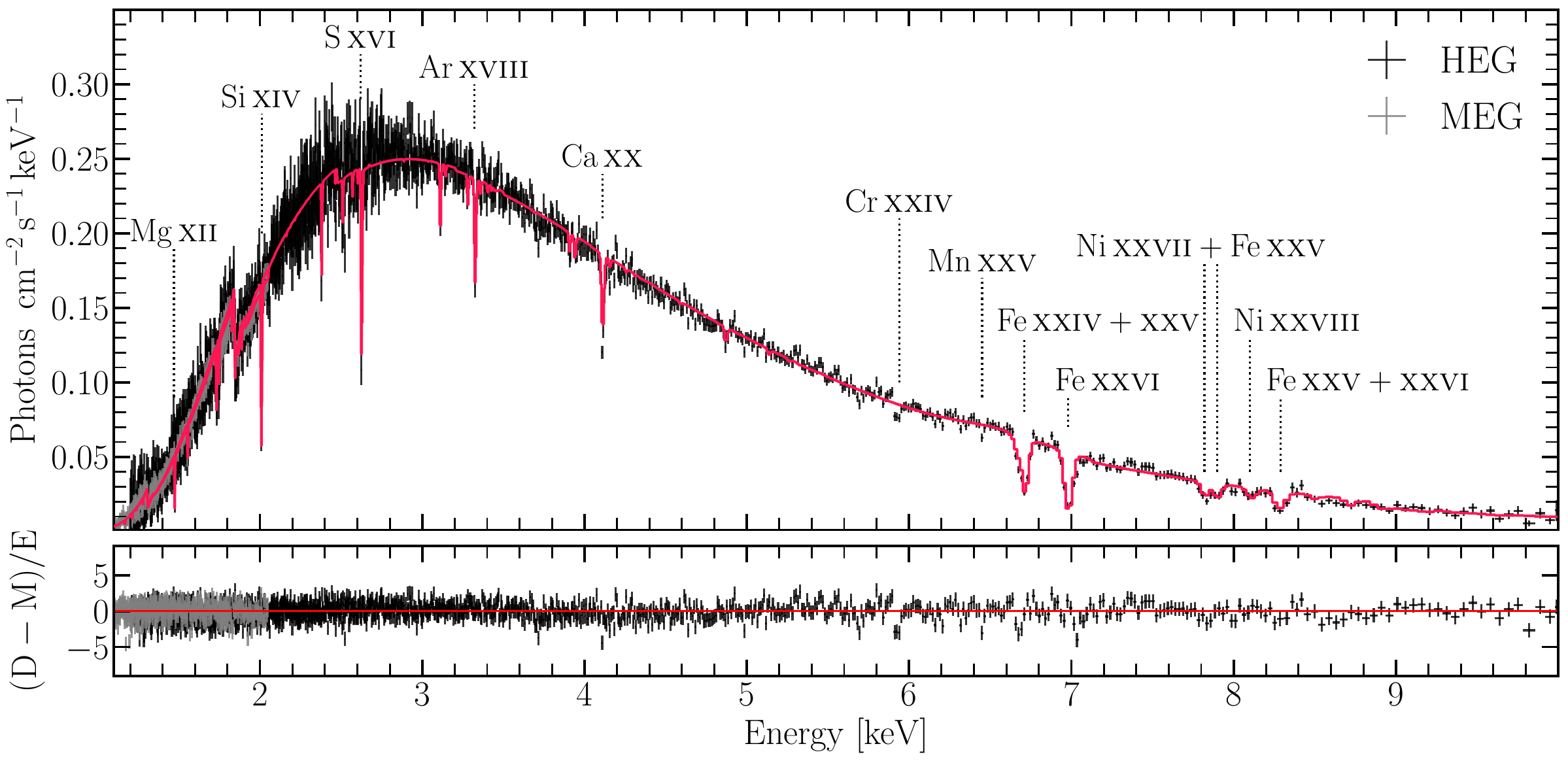}
\caption{HETG spectrum of observation 20294, epoch number 5. The obtained best fit is shown in magenta. HEG (black data points) and MEG (gray data points) spectra are fitted simultaneously. The residuals of the best fit are shown in the lower panel as $\rm (data-model)/error$. The respective HETG spectra of Epoch 1, 2, 4 are shown in Appendix \ref{sec:appendix}, Figure \ref{fig:all_hetg}.}
\label{fig:hetg}
\end{figure*}

\begin{figure*}[t]
\includegraphics[width=\textwidth]{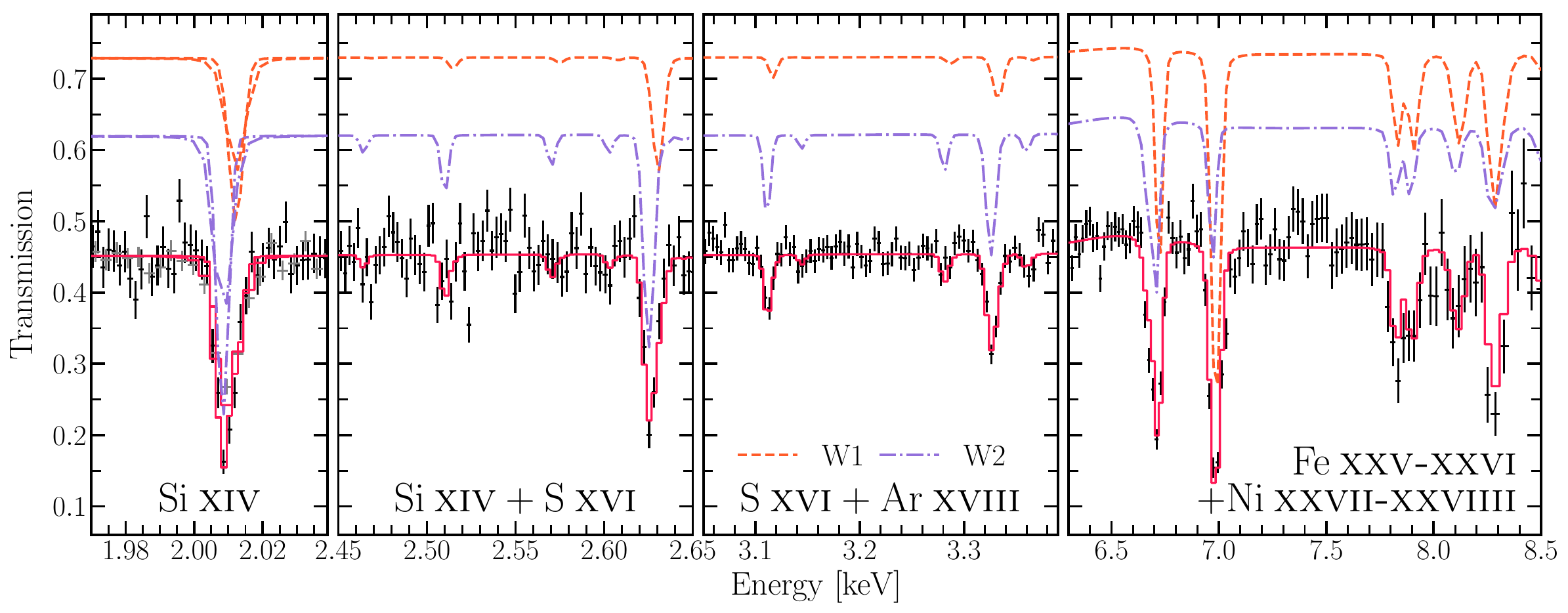}
\caption{Relative absorption contribution of the two wind components to the formation of primary resonance absorption lines detected in the \chandra/HETG observation during epoch 5. The transmission spectrum of the higher-ionization wind component is shown as an orange dashed line, while that of the lower ionization wind component is represented by a violet dot-dashed line. Transmission is defined here as the observed spectrum divided by the unabsorbed continuum.}
\label{fig:transmission}
\end{figure*}

The broad iron emission line in the complex Fe K band (6--8~keV) is detected in both \nicer and HETG data. We employed a Gaussian line (with full-width half maximum between 0.6 to 1.1~keV) to characterize this broad spectral feature. In addition to this narrower gaussian Fe line, we observed an improvement in the goodness of fit for the first three observations upon adding a very broad ($FWHM > 3\:\rm keV$) Gaussian line with $E_0\sim 5.8\ \rm keV$ to the continuum model. However, this new component exhibits weak degeneracy with the \texttt{bb} parameters. Therefore, we adopted a simpler model excluding the very broad line, which adequately describes the underlying emission. We also tested for a high-energy power-law component, as in previous studies \citep{Allen18}, by adding the \texttt{pow} component to our broadband model with low and high energy cutoffs to ensure correct energy balance in \spex. However, this did not significantly improve the fit, so we excluded the power-law component. Given the \nicer and HETG energy bands up to 10 keV, detecting and constraining this hard component is challenging. The final \spex model used to fit the continuum and the wind features: 
$$ {\tt (bb+mbb+gaus)*pion_2*pion_1*hot*amol}$$
where the two pion components are used to model the photoionisation absorption lines imprinted by the accretion disk (see Section \ref{sec:wind}). The best fit parameters are shown in Table \ref{tab:results}. We also estimated the observed and intrinsic (i.e. unabsorbed) flux assuming a distance of 7 kpc \citep{Bandyopadhyay99}.

\subsection{Accretion disk wind} 
\label{sec:wind}
The HETG spectra of GX 13+1 clearly show strong blueshifted H-like (Mg, Si, S, Fe, Ni) and He-like (Fe, Ni) absorption lines suggesting the presence of an outflowing ionized disc wind (see Figure \ref{fig:hetg} and \ref{fig:all_hetg}). Using the continuum described earlier and adding the \pion model, we recovered several wind properties, including its column density, ionization parameter $\log \xi$, line-of-sight outflow velocity \vout, and velocity width \vturb. The velocity width might be due to turbulent motions within the outflow, due to centrifugal flows or due to time variable systematic velocity. To describe all absorption lines accurately, a two-component outflow is required. The best-fit parameters are shown in Table \ref{tab:results}. The first wind component (W1) with a lower ionisation parameter ($\log \xi \sim 3.8 - 4.0$) and outflow velocity ($\vout \sim 300-500\rm\: \kms$) describes the \fexxv and \fexxvi lines together with the absorption features by lighter elements. The significant residuals on the high-energy wing of the \fexxvi line indicates the presence of a second component (W2) with a higher ionisation parameter $\log \xi \sim 4.3-4.7$) and systematic velocity ($\vout \sim 400-1200\rm\: \kms$). The uncertainties on the parameters of this high ionisation wind are larger because of its lower transmittance and known degeneracies between some of the photoionisation model parameters (e.g. $\xi$ and \NH). Figure \ref{fig:transmission} shows the two components' contributions to shaping the main resonance absorption lines detected in the HETG spectra of \gx.\\
We also considered a possible third component for the disk wind in \gx by adding another \pion component and performing the fit and the error search again. We explored a range of different \logxierg values between 2 and 5. However, no significant detection was found in any of the HETG spectra, with fit improvements over the two-component model of $\Delta C{\rm stat} < 10$. \\
During Epoch 3, observed solely with \nicer, \gx was in the horizontal branch. Following a similar procedure, we fitted the broadband continuum and photoionized wind, fixing the interstellar dust and neutral medium columns to the average values from the four \chandra/HETG observations. The disk wind is distinctly detected in this brief \nicer observation (see Figure \ref{fig:nicer}). Given the lower spectral resolution of \nicer, a single photoionization zone with $\log \xi = 4.86$ and $\NH = 3.4\times 10^{24}\ \rm cm^{-2}$ sufficiently accounts for the Fe K band absorption features. The wind velocities are constrained to the average values observed in the \chandra/HETG data. Further details on the best-fit model are presented in Appendix \ref{sec:appendix}.
 
\subsection{Jet} \label{sec:jet}
Radio emission was detected with the \vla during all four \chandra observation epochs. The resulting radio light curves are presented in the top panel of Figure \ref{fig:lc} and in Figure \ref{fig:vla}. The lowest flux densities were recorded during epoch 2, with average values $\left<f_\nu\right>$ of 0.37, 0.32, 0.24, and 0.21 mJy bm$^{-1}$ at frequencies of 5.25, 7.45, 14.4, and 17.2 GHz, respectively. In contrast, during epochs 1, 4, and 5, flux densities were approximately ten times higher across all frequencies. Notable variability was observed within each 5-hour epoch, with the most pronounced changes during epoch 5. Here, the radio flux density increased by a factor of about 5 over a period of 100 minutes at all four frequencies (see in Appendix \ref{sec:appendix_vla}, Figure \ref{fig:vla}). As expected for relativistic jets, higher frequencies, which are likely emitted from smaller regions, displayed finer substructure on $\approx$20-30 minute timescales, whereas the lower frequency curves appeared smoother. Perhaps unexpectedly, the lower frequency light curves peak at higher flux densities than high frequency light curves,  which we discuss in Section \ref{sec:disc:radio}.  

The radio spectra were steep overall, with average spectral indices of $\left<\alpha\right>=-0.49\pm0.01$, $-0.48\pm0.07$, and $-0.57\pm0.02$ for epochs 1, 2, and 4, respectively. During epoch 5, the spectra were steep both before and at the flare peak, but they transitioned to a flatter or inverted profile during the flare transition, with a maximum observed index of $\alpha \approx 0.14\pm0.15$. A similar, albeit minor, shift in spectral index was observed at the end of epoch 2, where $\alpha$ increased from approximately $-0.48$ to $-0.15$ as the flux density began to rise (Appendix \ref{sec:appendix_vla}, Figure \ref{fig:vla}).

\subsection{Time-resolved X-ray spectroscopy of Epoch 5} \label{sec:epoch5}

\begin{figure}[ht]
\includegraphics[width=1.05\columnwidth]{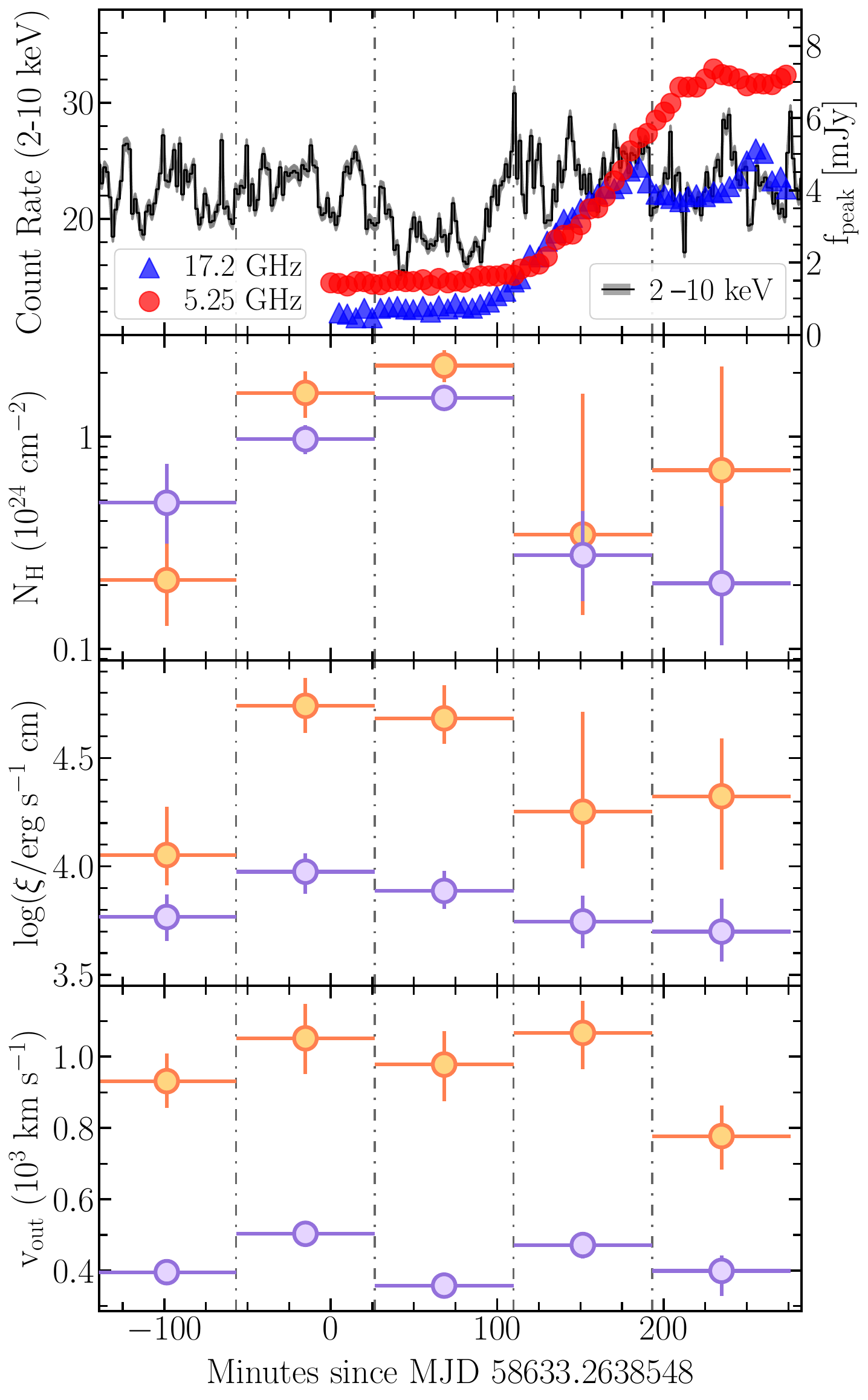}
\caption{Time-resolved spectral analysis of the accretion disk wind during Epoch 5. The top panel shows the X-ray light curve from \chandra/HETG (in black), alongside radio light curves at 5 and 17 GHz (represented by red circles and blue triangles, respectively) from \vla observations. Vertical dotted-dashed lines divide the \chandra/HETG data into five intervals of 5 ks each, for which we extracted and modeled the HETG spectra with a focus on the two wind components. The bottom three panels display the optimal values of wind parameters, with the higher-ionization wind component shown in orange and the lower-ionization component in violet.}
\label{fig:epoch5}
\end{figure}

%
\begin{table*}
\caption{Time-resolved spectral analysis of the two wind components for Epoch 5. The turbulence velocity is fixed to the averaged value of $\vout=40\: \kms$.}
\label{tab:time_resolve}
\begin{threeparttable}        
\begin{tabular}{c c c | c c c c c } 
\hline\hline       
\noalign{\vskip 0.15mm}
Comp. & Par. & Units & Interval 1 & Interval 2 & Interval 3 & Interval 4 & Interval 5 \\
\noalign{\vskip 0.15mm}
\hline
\noalign{\vskip 0.15mm}
\multirow{4}{*}{\texttt{pionW1}} & $N_{\rm H}$ & $10^{23}\:\rm cm^{-2}$ & $4.9_{-1.7}^{+2.6}$ & $9.7_{-1.4}^{+1.6}$ & $15.2_{-2.0}^{+2.2}$ & $2.8_{-1.0}^{+1.6}$ & $2.0_{-1.0}^{+2.6}$ \\
& \multicolumn{2}{c|}{\logxierg}  & $3.77\pm0.10$ & $3.97_{-0.10}^{+0.09}$ & $3.88\pm0.09$  & $3.74\pm0.12$ & $3.70\pm0.15$ \\
& \vout & \kms & $400\pm30$ & $500_{-40}^{+30}$ & $360\pm30$  & $470_{-40}^{+30}$ & $400_{-70}^{+40}$ \\
\noalign{\vskip 0.15mm}
\hline
\noalign{\vskip 0.15mm}  
\multirow{4}{*}{\texttt{pionW2}} & $N_{\rm H}$ & $10^{23}\:\rm cm^{-2}$ & $2.1_{-0.8}^{+2.5}$ 
& $16.1_{-3.7}^{+4.1}$ & $21.6_{-3.5}^{+4.0}$ & $ 3.5_{-2.0}^{+12.5}$ & $ 6.9_{-4.8}^{+14.5}$ \\
& \multicolumn{2}{c|}{\logxierg}  & $4.05_{-0.15}^{+0.22}$ & $4.74\pm0.12$ & $4.68_{-0.12}^{+0.15}$ & $4.25_{-0.26}^{+0.46}$ 
& $4.32_{-0.34}^{+0.27}$ \\
& \vout & \kms & $930\pm80$ & $1050\pm100$ & $980\pm100$  & $1070_{-100}^{+90}$ & $780\pm90$ \\

\noalign{\vskip 0.15mm}
\hline
\end{tabular}
\end{threeparttable}
\end{table*}

During epoch 5, the \vla recorded a significant increase in the radio flux density of \gx, by approximately a factor of 5 over 6-7 ks ($\sim1.5\:\mbox{-}\:2$ hours), varying with radio frequency. The X-ray light curve, meanwhile, exhibited irregular variability, with a twofold rise over $2~\rm ks$ (0.5 hours) preceding the increase in radio flux density (refer to the top panel of Figure \ref{fig:epoch5}). We explored potential cross-correlations between the radio and X-ray light curves using the PYCCF.py package\footnote{\url{https://bitbucket.org/cgrier/python_ccf_code/src/master/}} \citep{Peterson98, Sun18}. The cross-correlation function was flat and low, suggesting a weak or absent relationship between the radio and X-ray light curves.

To further probe the relationship between radio emission and wind properties, potentially linking the jet and disk wind, we divided the \chandra observation for epoch 5 into five intervals of approximately 5 ks each. This partitioning aligned interval $i_3$ with the radio low state, $i_4$ with the transition phase, and $i_5$ with the high radio state.
Following the data reduction approach in Section \ref{sec:data_reduction_chandra}, we extracted HETG spectra (first orders of HEG and MEG) for each interval, grouped the data, and performed fits over a narrower energy range of 1.2-7.5~keV. The model from the complete 20194 \chandra/HETG observation was adopted as the base, with the hydrogen column density of the cold ISM, dust column density, and the average line energy and FWHM of the Gaussian model fixed. In contrast, the parameters of the two blackbody components (\texttt{bb} and \texttt{mbb}) and Gaussian normalization were allowed to vary to account for continuum changes.
Finally, we determined the free parameters ($N_{H}$, \logxi, \vturb, and \vout) for the two \pion components in each interval. Figure \ref{fig:epoch5} presents the best-fit parameter values with uncertainties alongside the X-ray and radio light curves. The column density of the low-ionisation component seems to decrease as the radio flux increases. The other parameters do not show any strong evolution. Additionally, we examined the evolution of the continuum model parameters, such as the temperatures of \texttt{mbb} and \texttt{bb}, and found no significant correlation with radio emission.
Finally, we determined the free parameters ($N_{H}$, \logxi, \vturb, and \vout) for the two \pion components in each interval. Figure \ref{fig:epoch5} and Table \ref{tab:time_resolve} present the best-fit parameter values with uncertainties alongside the X-ray and radio light curves. The column density of the low-ionisation component seems to decrease as the radio flux increases. The column density values between intervals $i_3$ and $i_4$ differ by more than $3\sigma$. The column densities of the highly ionised components also decrease, although the change is not statistically significant due to large uncertainties. The other parameters do not show any strong evolution. Additionally, we examined the evolution of the continuum model parameters, such as the temperatures of \texttt{mbb} and \texttt{bb}, and found no significant correlation with radio emission.

\section{Discussion} \label{sec:discussion}
Our simultaneous radio and (high-resolution) X-ray observations provide new insights into the interplay between the accretion disc wind, accretion flow state, and radio jet outflow in \gx. Here, we discuss the key findings of this multi-wavelength campaign, exploring the relationship between the accretion disc wind and relativistic jet outflow in this near-Eddington neutron star X-ray binary. We compare our results with previous studies on black hole X-ray binaries and low-Eddington neutron star X-ray binaries to build a broader context.

\subsection{Continuum and flaring branch observation}

We modeled the X-ray continuum using two thermal emission components, denoted as \texttt{bb} and \texttt{mbb}. The temperatures of these components, $T_{\rm bb}$ and $T_{\rm mbb}$, range from 1.9 to 2.2 keV and 1.65 to 2.05 keV, respectively, consistent with previous broadband spectral studies \cite[e.g.,][]{Allen18}. We included a Gaussian line to account for the broad emission feature associated with the Fe K$\alpha$ line, previously detected in observations with \asca, \xmm, \xte, and \chandra \citep[e.g.,][]{Asai00, Ueda01, Sidoli02, Homan04, Ueda04, DiazTrigo12}. The average energy of this line, $E_0$, lies between 6.44 and 6.59 keV, while its full width at half maximum (FWHM)\footnote{Past studies often report the Gaussian width, $\sigma$, which relates to the FWHM by $\rm FWHM = 2.3548\,\sigma$} spans from 0.59 to 1.1 keV. These values are in agreement with previous measurements, such as those from earlier HETG spectra \citep{Allen18}. The column density of the neutral ISM varies across the observations, suggesting some level of degeneracy between the continuum components (\texttt{bb} and \texttt{mbb}), the dust absorption model (\texttt{amol}), and the Galactic absorption (\texttt{hot}). Nevertheless, the values are consistent with the \NH\ measurements reported in previous works \citep[e.g.,][]{Ueda04, DiazTrigo12, Saavedra23}, particularly with \cite{Zeegers19}, who carried out a detailed study of the foreground Galactic absorption. \\

This campaign marks the first \chandra observations of \gx on its flaring branch, observed in epochs 1, 4, and 5 (see Figure \ref{fig:hid}). Previously, \chandra observed the source on its normal and horizontal branches only \citep{Homan16}. Combined with archival observations, these data now cover GX 13+1’s entire Z-track, enabling us to track the wind properties across all branches. 

\begin{figure*}[t]
\includegraphics[width=\textwidth]{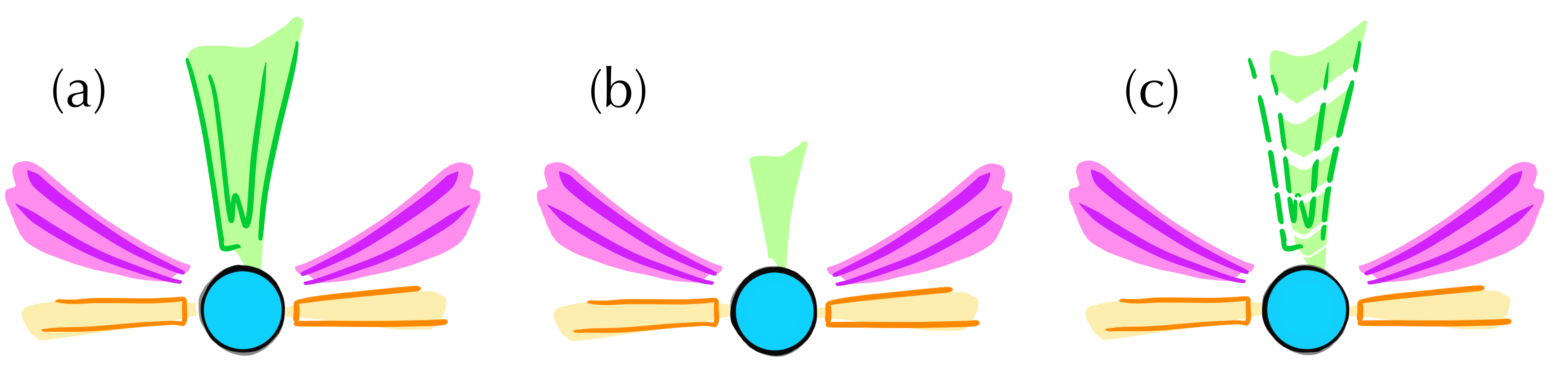}
\caption{Illustrative overview of the interaction between the persistent accretion disc wind (in magenta/violet) and the evolving radio jet outflow (in green) in the accreting neutron star (in blue) X-ray binary \gx. \textit{Case (a)}: A strong wind and active radio jet are observed during Epochs 1 and 4, corresponding to the flaring branch state. \textit{Case (b)}: During Epoch 2, as the source transitions to its normal branch, the radio jet notably weakens while the winds remain robust. \textit{Case (c)}: During Epoch 5, \gx exhibits an intensified radio jet on its flaring branch, with no substantial changes in the accretion disc wind.}
\label{fig:wind_jet}
\end{figure*}

\subsection{Wind properties and variability}

\cite{Allen18} conducted an extensive study of the wind properties of \gx, analyzing seven \chandra/HETG spectra taken from 2002 to 2011, totaling 180 ks of observation. They identified two variable wind components necessary to describe the narrow absorption features. The lower-ionization component exhibited \NH values from $0.3$ to $0.8 \times 10^{24} \: \rm cm^{-2}$, log $\xi$ values between $2.71$ and $2.98$, turbulent velocities of $60$ to 380 km/s, and outflow velocities of $310$ to 790 km/s. The higher-ionization component had \NH values from $1.2$ to $2.2 \times 10^{24} \: \rm cm^{-2}$, log $\xi$ values from $3.99$ to $4.21$, $v_{\text{turb}}$ between $380$ and $820 \: \text{km/s}$, and $v_{\text{out}}$ from $760$ to $1170 \: \text{km/s}$. Generally, the wind properties are consistent with the value ranges observed between 2018 and 2019. The main significant difference is the lower ionization parameters observed in the earlier \chandra observations by \cite{Allen18}. Since we detected the same absorption lines (e.g., \ion{Mg}{xii}, \ion{Si}{xiv}, \ion{S}{xvi}, \ion{Ar}{xviii}, \ion{Ca}{xx}, \ion{Fe}{xxv}, and \ion{Fe}{xxvi}), we suspect that the observed variation is due to differences in SED modeling. The ionization parameter depends on the ionizing luminosity, which is derived from the broadband SED. Furthermore, as \cite{Mehdipour16} demonstrated, the choice of atomic codes and photoionization models can systematically affect ionization parameter estimates. Unlike \cite{Allen18}, who used XSTAR-generated photoionization tables, this analysis employed two \pion models to characterize the wind. \\
Additionally, the turbulent velocities in both wind components are systematically lower than those reported by \cite{Allen18}, potentially explaining the higher column densities obtained here. However, as illustrated in Figure \ref{fig:transmission}, the strongest ionisation lines are well-fitted with narrow line profiles, with no evidence of additional broadening in the current HETG observations. It is also worth noting that the turbulence velocity measurements by \cite{Allen18} have substantial uncertainties, and most of their results are consistent with the present findings within a 2$\sigma$ confidence level.

Over nearly two decades of monitoring with \chandra and \xmm, the absorption features display no substantial variability \citep{Ueda04,DiazTrigo12,Allen18}. The two components modelling these features maintain similar ionization parameters and column densities suggesting a persistent, relatively smooth outflow rather than large, isolated clumps of ionized material \citep[see also][]{Allen18}. Additionally, wind properties show minimal variation along the Z-track (see Section \ref{sec:wind}). \\

We cannot rule out the possibility that Fe \textsc{xxv} and \textsc{xxvi} K$\alpha$ lines may be heavily saturated \citep{Tomaru20} providing an incorrect measurement of the column density of the wind components. For high-column densities, line depth does not increase below zero intensity, so raising the ion absorption columns beyond this point only slightly broadens the line, making it difficult to pinpoint the ion column. However, we detect Fe K$\beta$ lines and K$\alpha$ absorption lines from elements like argon and calcium, which are less likely to be saturated, providing clearer constraints on the total outflow column density under the assumption of solar abundances.

\subsection{The Origin of Radio Emission}
\label{sec:disc:radio}
 The observed radio emission generally has a steep spectrum, which is intriguing. Steep radio spectra may argue against a compact, partially self-absorbed synchrotron jet \citep{Blandford79}, which is typically associated with hard-state LMXBs \citep{Fender01, Corbel02}. Instead, steep radio spectra are more characteristic of discrete, `ballistic' ejecta \citep{Mirabel94, Bright20}. However, such ejecta would typically fade over time and are expected to be transient, associated with X-ray spectral state transitions \citep{Mirabel94, Hjellming95}, which also appears inconsistent with the behavior of \gx. Nevertheless, the flatter radio spectra observed during the flare in Epoch 5 resemble an expanding synchrotron plasma transitioning from optically thick to optically thin at the flare peak, with the caveat that in this type of expansion, higher-frequency radio emission typically peaks at higher flux densities \citep{vanderlaan66}. The opposite is observed in \gx during Epoch 5. 

The steep radio spectrum observed during non-flaring periods remains puzzling but may indicate a variable jet superimposed on a more persistent radio source. The emission cannot originate from large-scale regions, as such emission would be filtered out due to the \vla data being taken in A and B configurations. Consequently, the source must be compact, and variability is to be expected. The observed month-long variability could be attributed to adiabatic expansion losses in the emitting region, which is subsequently replenished. The apparent lack of interaction between the wind and the jet suggests that the wind's stability does not significantly impact the jet's variability, allowing the jet to fluctuate independently.  

The steady nature of the X-ray wind suggests it is unlikely to produce shocks in the surrounding interstellar medium. However, during the normal branch observation—when the radio flux density is at its lowest level—the wind exhibits the smallest outflow velocities (280~km/s and 370~km/s for wind components 1 and 2, compared to average velocities of 450~km/s and 1100~km/s in other observations). We investigated the potential radio emission from the two-component wind observed in \gx. To estimate the maximum mass outflow rate of the wind, we used Equation 7 from \citet{Kosec20}, assuming an isotropic geometry, a full volume-filling factor ($C_{v}=1$), and a complete wind launch solid angle ($\Omega=4\pi$). We then converted the estimated mass outflow rate ($\dot{M}_{\rm out} \simeq 6\times 10^{-6} \rm~M_{\odot}/year $) into a predicted radio flux density using Equation 6 from \cite{Blustin09}, maintaining similar assumptions and adopting a source distance of 7~kpc. This yielded a predicted radio flux density of $S_{17~\rm GHz} \simeq 0.05~ \rm mJy$, whereas the observed 17~GHz radio flux density in this campaign ranges between 0.2 and 4~mJy. Thus, the wind can account for only a small fraction of the observed radio emission. Moreover, since the radio flux density $S_{\nu}$ scales with the outflow velocity as $S_{\nu} \propto (\dot{M}_{\rm out})^{4/3} \propto (v_{\rm out})^{4/3}$, a factor of 6 change in outflow velocity would be required to explain the radio flux variability between Epoch 1 and Epoch 2. However, the observed velocity changes are only factors of 2–3, making it unlikely that wind variability alone is responsible for the radio flux changes.  

\subsection{The Wind the Jet and X-ray State Coupling} \label{sec:wind_disk_jet}
We discuss here the relations between the ionised wind, the radio jet and the X-ray state that we observed in this campaign comparing with what has been observed in previous observations and other X-ray binaries. 
\subsubsection{The Wind and the X-ray State}
All \chandra observations in this study show strong, blueshifted absorption lines from a high-ionization accretion disk wind, consistent with archival data. Additionally, this wind is detected in all archival \xmm/EPIC-pn \citep[e.g.][]{Sidoli02,DiazTrigo12} and current \nicer X-ray spectra, identified by \ion{Fe}{xxv} and \ion{Fe}{xxvi} resonance absorption lines in the Fe K energy band. This persistent wind shows no strong correlation with the Z-track position, suggesting that the mass accretion rate may not vary strongly along the Z-track. This hypothesis aligns with previous work \citep[e.g.,][]{DiSalvo00,Homan02,Homan10}. For example, \cite{Lin09} analyzed the spectral evolution of XTE J1701-462 during its 2006 outburst and concluded that the mass accretion rate stayed constant along the Z track. The mass accretion rate appears to drive the secular evolution of the Z-track, while movement along the Z branches seems to be governed by other mechanisms (e.g., shrinking inner disk radius, increased Comptonization of the disk emission, or an increase in the boundary layer's apparent size) that operate at roughly constant accretion rates.

\subsubsection{The Jet and the X-ray State}

Simultaneous X-ray and radio observations of Z sources have revealed a connection between their spectral branches and radio brightness. \citet{Penninx88} first demonstrated this relationship in GX~17+2, showing that the radio emission varied systematically with the position in the X-ray hardness-intensity diagram (HID). Specifically, the radio brightness increased as the X-ray spectral state transitioned from the flaring/normal branch to the horizontal branch, where the strongest radio emission was observed. This behavior, initially reported in GX~17+2, has also been identified in Cyg~X-2 \citep{Hjellming90a} and Sco~X-1 \citep{Hjellming90b}. In both Cyg~X-2 and GX~17+2, the coupling between radio and X-ray states occurs on short timescales ranging from tens of minutes to a few hours. \citet{Migliari06} proposed a model to explain this coupling, suggesting that the compact object is primarily responsible for the strong radio emission in the horizontal branch and, to a lesser extent, in the normal branch. They further hypothesized that transitions between X-ray spectral states are linked to changes in jet emission, with transient optically thin radio flares occurring at the horizontal-to-normal branch transition.

In early simultaneous X-ray and radio observations of \gx, \citet{Homan04} reported a strong dependence of radio brightness on the X-ray state. Specifically, they observed significant jet activity during the normal and horizontal branches (hard spectral state), accompanied by weak or negligible radio emission during the flaring branch (soft spectral state). In their study, radio flux densities during the hard state were 4 to 18 times higher than the maximum values observed in the soft state.

\begin{figure*}[t]
\includegraphics[width=0.495\textwidth]{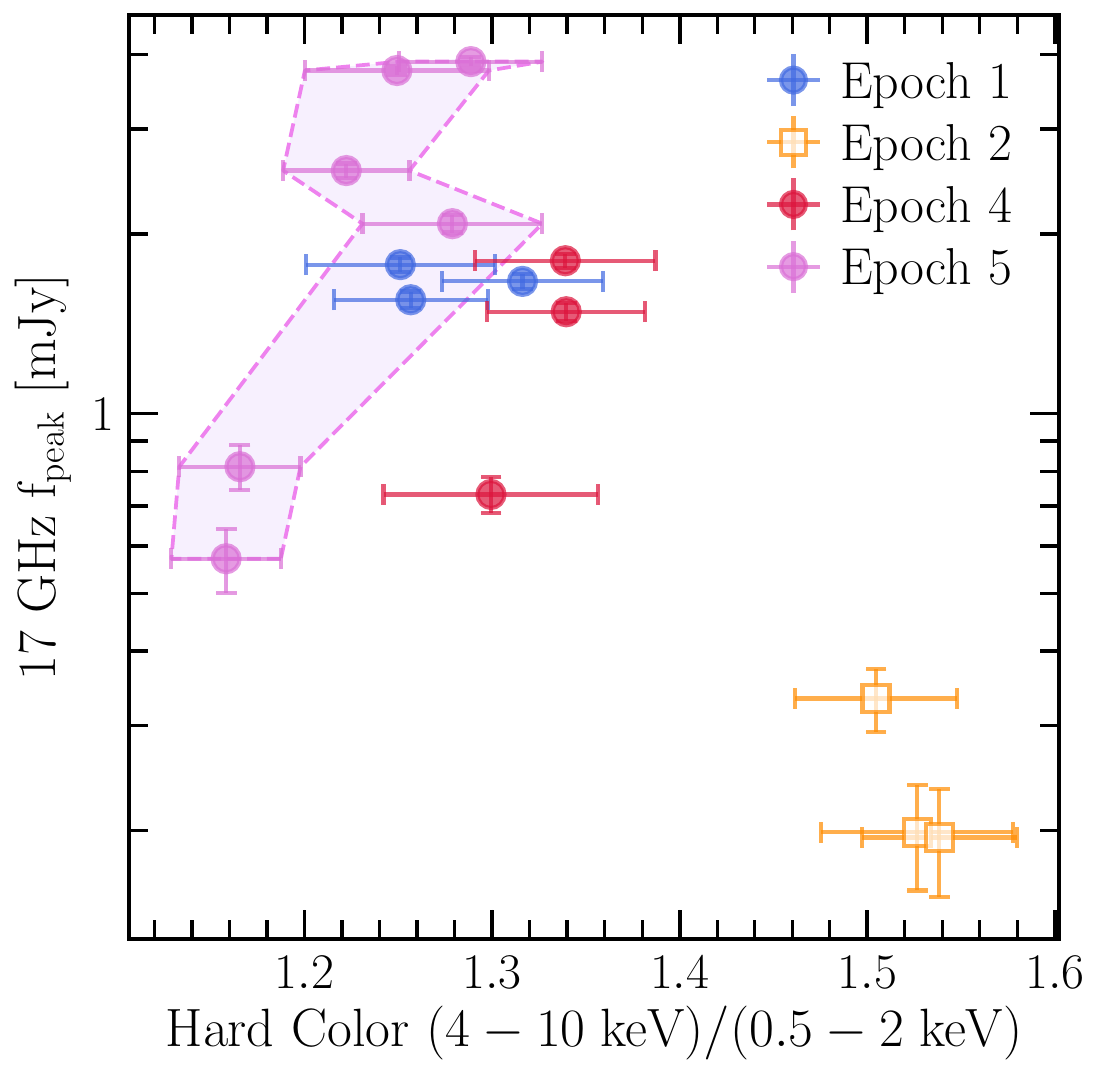}
\includegraphics[width=0.495\textwidth]{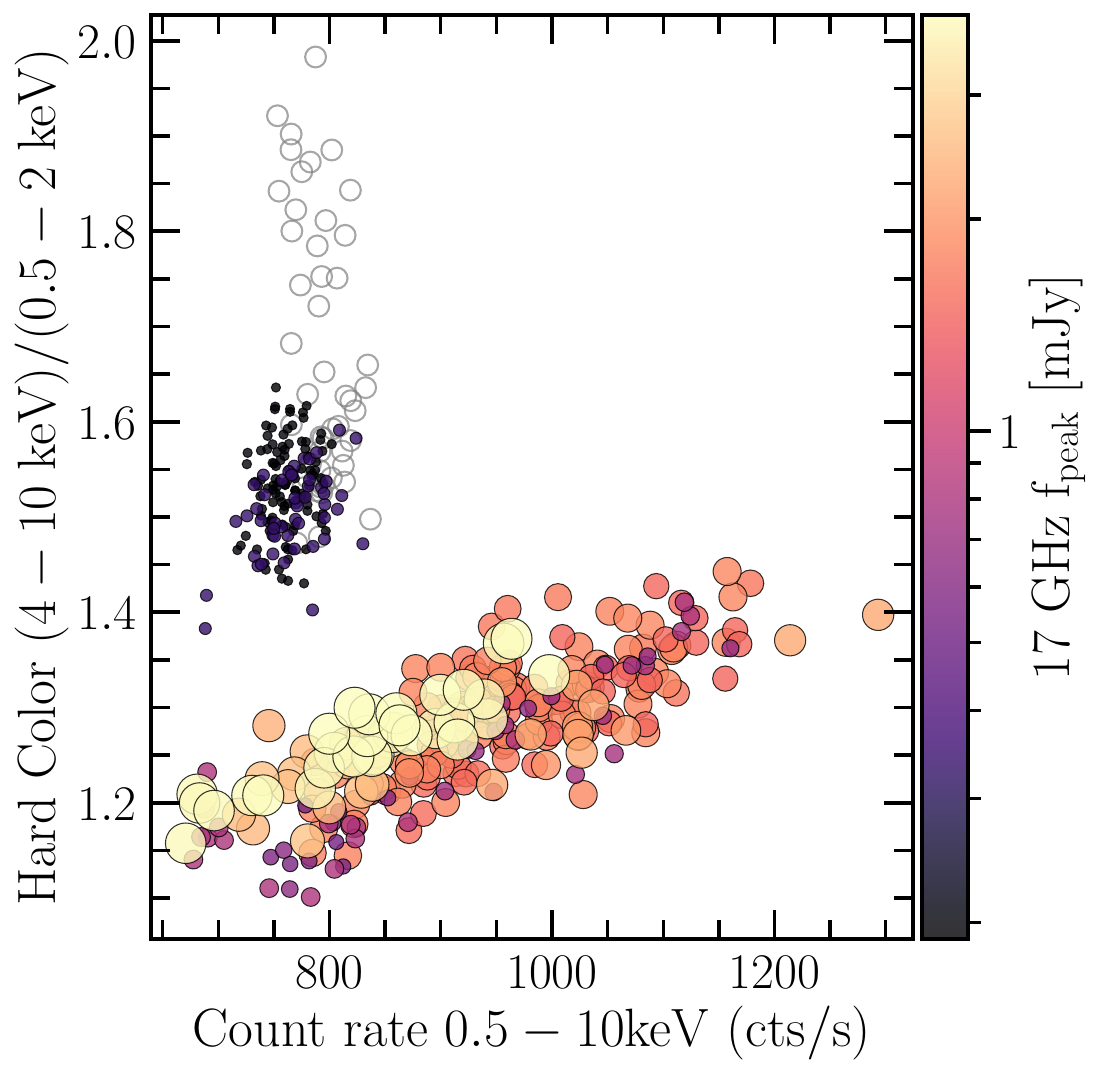}
\caption{\textit{Left}: Radio flux densities at 17 GHz plotted against the X-ray hard color for the 2018/2019 simultaneous \nicer/\vla observations. Solid circles represent flaring branch observations (Epochs 1, 4, and 5), which show higher radio flux densities, while empty squares correspond to normal branch observations (Epoch 2) characterized by weaker radio flux. The shaded pink region between the dashed lines illustrates the radio-hard color evolution during the pronounced and rapid increase in radio flux density observed in Epoch 5. \textit{Right}: Hardness-intensity diagram of the \nicer observations strictly simultaneous with the \vla data. The size and color of the circles represent their interpolated 17 GHz radio flux density values. For reference, empty grey circles indicate the \nicer observations on the horizontal branch during epoch 3 where we do not have radio coverage. 
}
\label{fig:radio_color}
\end{figure*}

In contrast, our campaign reveals different behavior. Maximum radio emission was observed during all three flaring branch observations (Epochs 1, 4, and 5). Conversely, the radio flux density during the normal branch observation was approximately 5 to 20 times lower than the peak values recorded during the flaring branch. The left panel of Figure~\ref{fig:radio_color} displays the 17 GHz radio flux density versus the hard color used in the HID. For this analysis, X-ray data were binned per intra-observation NICER snapshot, and the corresponding radio flux density values were interpolated. The data points reveal an apparent anti-correlation between radio flux density and hard color, which challenges the typical behavior observed in other Z sources such as GX~17+2, Cyg~X-2, Sco~X-1, and XTE~J1701-462. For example, simultaneous X-ray and radio observations of GX~17+2 consistently showed higher radio flux densities in harder X-ray states (horizontal and normal branches) \citep{Penninx88, Migliari07}. Unfortunately, we lack simultaneous X-ray and radio data during the horizontal branch of \gx, limiting a direct comparison with other Z sources (right panel of Figure \ref{fig:radio_color}).

Figure~\ref{fig:radio_color} also highlights the evolution of radio flux density and hard color during Epoch 5 (left panel). During this epoch, the X-ray spectrum hardened as the radio flux density increased. This contrasts with GX~17+2, which exhibited no significant change in hard color during a radio flare observed with the VLA in 2002 \citep{Migliari07}. Additionally, transient radio flares in Z sources are typically associated with horizontal-to-normal branch transitions \citep{Migliari06}. However, in our campaign, \gx exhibited a significant increase in radio flux density during Epoch 5, when the source was in the flaring branch. The radio flux density began at a low-to-intermediate level and increased by a factor of approximately 5 over 100 minutes. During this transition, the spectral index shifted from a median value of $-0.6$ to a flat spectrum ($\alpha \sim 0$ to $0.2$), before returning to a median of $-0.5$ as the flux stabilized at higher levels.

This behavior highlights the unique nature of \gx. While the source exhibits high radio emission typical of Z sources, the lack of a clear correlation between the radio flux and the spectral branches remains puzzling. Specifically, it is unclear why \gx displayed the expected Z-source behavior in earlier \xte/\vla observations from 1999—characterized by strong radio emission in the harder normal/horizontal branches and weak radio emission in the softer flaring branch \citep{Homan04, Homan16}—yet shows the opposite pattern in the current campaign. Notably, the radio and X-ray properties of \gx, including the mean and maximum flux levels, the HID tracks, and variability, have not changed significantly between the two epochs. This consistency makes it challenging to reconcile the contrasting jet and X-ray state coupling observed in these two sets of observations.

\subsubsection{The Wind and the Jet}
\gx does not exhibit the jet-wind anticorrelation commonly observed in high-inclination, typically less luminous black hole X-ray binaries. In these systems, equatorial winds are typically present in jet-free soft states and absent in hard states where a strong jet is detected \citep{Neilsen09, Ponti12a, Miller12}. Similar correlations between accretion disk winds and spectral hardness have been observed in low-Eddington neutron-star LMXBs such as EXO 0748-676 and AX J1745.6-2901 \citep{Ponti14, Ponti15}. In both cases, no significant absorption lines from the wind are detected in their hard-state spectra. In the black hole X-ray binary GRS 1915+105, the wind drives approximately the same mass loss as the radio jet, suggesting the system maintains a balanced equilibrium between mass accretion and outflow, regardless of its spectral state or outflow mechanisms \citep{Neilsen09}. In contrast, \gx shows simultaneous presence of both wind and jet, indicating they are not mutually exclusive. The two outflows appear to operate independently and not competing for the same energy reservoir.

\gx is highly luminous, with near-Eddington X-ray luminosity, $L_{\rm X}\gtrsim 0.5L_{\rm Edd}$ \citep[e.g.,][]{Allen18}. Due to the high Eddington fraction, radiation pressure may play a significant role in wind launching, potentially explaining \gx's persistent wind, which seems to flow independently of the accretion disk state and radio jet. However, radiation pressure alone is inefficient at transferring momentum to a highly ionized wind \citep{Tomaru20}. X-ray heating and thermal pressure or strong magnetic fields may provide the additional force needed to accelerate the hot wind (see Section \ref{sec:wind_variability} for more details). 

In time-resolved X-ray spectroscopy of Epoch 5, no significant correlation between wind properties and radio jet power was observed. However, there is an intriguing decrease in wind column densities alongside an increase in radio flux, particularly in the low-ionization component where uncertainties are smaller. This behavior might resemble the anticorrelation between wind and jet reported in the BH X-ray binary sample by \cite{Ponti12a}. The key distinction in \gx is that the wind remains detectable and relatively strong even when radio emission is high, with only a marginal decrease in column density. On a larger scale, this behavior might parallel observations in radio-loud galaxies. \cite{Mehdipour19} identified an anticorrelation between the column density of ionized winds and radio-loud jet power in radio-loud AGN, suggesting that wind-jet bimodality could stem from the configuration of magnetic fields near the black hole. 

A similar phenomenon is observed in broad absorption line (BAL) quasars, where strong jet production events appear to suppress the high-ionization BAL winds typically present in weak jet states \citep[e.g.,][]{Shankar08, Reynolds17}. However, this conclusion should be approached with caution, as the magnetic-field configurations, which might play a critical role in launching outflows, differ significantly between neutron stars and supermassive black holes.

Thanks to optical and near-infrared (nIR) observations of black hole transients, the number of LMXBs with confirmed simultaneous jets and winds has rapidly increased. These observations have revealed cold winds in several black hole transients, at times when radio jets were also being launched. Examples of such sources are MAXI J1820+070 \citep{munoz-darias2019,bright2018}, MAXI J1348$–$630 \citep{Carotenuto2021,Panizo-Espinar2022}, GRS 1915+105 \citep{Sanchez2023}, and Swift J1727.8$-$1613 \citep{Miller-Jones2023,Mata2024}. These sources were likely all accreting at lower Eddington fractions than \gx. It is not clear how the optical/nIR winds in these sources are related to the X-ray winds that are predominantly detected in the softer X-ray states. We are not aware of simultaneous wind detections in X-rays and the optical/nIR that could shed light on this.

\subsection{Wind Launching Mechanism} \label{sec:wind_variability}
Determining the wind-driving force is key to understanding its physical connection with the accretion disk and radio jet. Various wind-driving mechanisms, including magnetic driving, Compton heating, and radiation pressure, are debated. The launching radius helps differentiate among them: magnetic driving is more likely at small launching radii \citep[e.g.,][]{Miller06c, Miller16, Neilsen16, Fukumura21}, while thermal driving is effective at large distances from the compact object, where the thermal velocity exceeds local escape velocity \citep[e.g.,][]{Proga02}. 

Assuming the wind velocity exceeds the escape velocity, $v_{\rm esc}$, we can estimate the lower limit of the launching radius using $R > 2GM/v_{\text{out}} \simeq 4\times10^{10}\:\rm cm$, where $G$ is the gravitational constant and $M$ is taken as the canonical neutron star mass, $M=1.4\:\rm M_{\odot}$. This limit is consistent with previous estimates of $R\approx10^{10-11}\rm \: cm$ \citep{Ueda04, DiazTrigo12, dAi14, Allen18}, though different assumptions were used. \cite{Ueda04} and \cite{DiazTrigo12} assumed that the slab thickness, $d$, of the ionized plasma is comparable to the distance ($0.1 \lesssim d/r \lesssim 1)$, while \cite{dAi14} and \cite{Allen18} fixed the electron density, $n$, at $10^{12}\:\rm cm^{-2}$. Studies of \gx indicate that the disc wind is likely driven by Compton heating, as the estimated launching radius exceeds $0.1\times R_{\rm C}$, where $R_{\rm C}$ is the Compton radius, equal to $8\times 10^{10}\:\rm cm$ for \gx \citep[e.g.,][]{Allen18}. A thermal wind can be efficiently accelerated just beyond $0.1 R_{\rm C}$ \citep[e.g.,][]{Begelman83a}. Alternatively, \cite{Ueda04} suggested a radiation-driven origin for the disk wind, proposing that radiation pressure could be effective even at sub-Eddington luminosities. Full radiation hydrodynamic simulations by \cite{Tomaru20} further indicate that a thermal-radiative mechanism is consistent with the absorption lines observed in the \chandra HETGS third-order spectra. At near-Eddington luminosities, radiation pressure on ions and electrons may add momentum to the wind, increasing its velocity.

While the growing evidence points toward a thermal-radiative origin for the wind in \gx, current data do not rule out a magnetically driven mechanism. Upcoming and future observations with high-resolution spectral instruments on \xrism \citep{Tashiro20,Tashiro24} and {\it NewAthena} \citep{Cruise25} will be crucial in identifying the nature of the accretion disk wind in low-mass X-ray binaries. Winds driven thermally, radiatively, and magnetically each produce distinct absorption line profiles that can be distinguished in microcalorimeter spectra \citep{Fukumura22, Gandhi22}. 

\section{Summary} \label{sec:conclusion}

\cite{Homan16} suggested the possible coexistence of accretion disc wind and radio jet outflow in X-ray binaries with near/high Eddington rates. To validate this hypothesis, we conducted an extensive, multi-wavelength campaign to observe \gx, a bright neutron star X-ray binary, to examine the relationship between mass accretion flow, ionized outflows, and jet emission. Over the course of a year, we performed observation at five epochs, four of which included simultaneous \vla, \chandra/HETG, and \nicer\ data, while one included \nicer\ data alone. Key findings from our analysis of the radio light curves and X-ray spectra are as follows:

\begin{itemize}
\item The \chandra/HETG observations of \gx\ covered the entire Z track, indicating a persistent wind across all branches. A two-component wind model, with outflow velocities between 300 and 1200 km/s and ionization parameters ranging from $\log \xi = 3.8$ to $\log \xi = 4.7$, successfully fit the dense array of absorption lines observed in the data.

\item Strictly simultaneous \chandra and \vla observations reveal that accretion disk X-ray winds and radio jets co-exist in \gx. The radio emission varies within and across the different \vla observations whereas the properties of the wind change only slightly. 

\item In this campaign, \gx does not exhibit the typical correlation between radio emission and X-ray state observed in Z-sources, where steady and strong jets are associated with the horizontal branch, weaken on the normal branch, and are suppressed on the flaring branch. Instead, we observe an anticorrelation between radio flux density and hardness color, with the radio jet being particularly strong during all flaring branch observations and variable in one of them, and significantly weaker when the source is on the normal branch.

\item The contrasting variability between the wind and radio jets suggests that these outflows likely originate from distinct mechanisms and do not share a common launching process. In this near-Eddington source, the strong radiation field may drive the persistent wind, accounting for its steady presence. 
\end{itemize}

\begin{acknowledgments}
We thank the referee for the thorough review of the manuscript and for the helpful insights that have improved the quality of the paper. DR is grateful Dr Peter Kosec and Dr Missagh Mehdipour for insightful discussions on wind modeling, as well as Dr Christos Panagiotou for his assistance in estimating the cross-correlation function between the radio and X-ray light curves. DR also acknowledges the valuable advice and mentorship provided by Prof. Erin Kara and Prof. Irina Zhuravleva. \\
During the project, D.R. received support from NASA through the Smithsonian Astrophysical Observatory (SAO) contract SV3-73016 to MIT for Support of the Chandra X-Ray Center (CXC) and Science Instruments, as well as the Margaret Burbidge Fellowship supported by the Brinson Foundation. J.H.\ acknowledges support from NASA through Chandra Award Number GO8-19028X issued by the CXC. The CXC is operated by the Smithsonian Astrophysical Observatory on behalf of NASA under contract NAS8-03060. 

\end{acknowledgments}

\appendix

\section{\chandra/HETG and \nicer spectra} \label{sec:appendix}
\begin{figure*}[t]
\centering
\includegraphics[width=.8\textwidth]{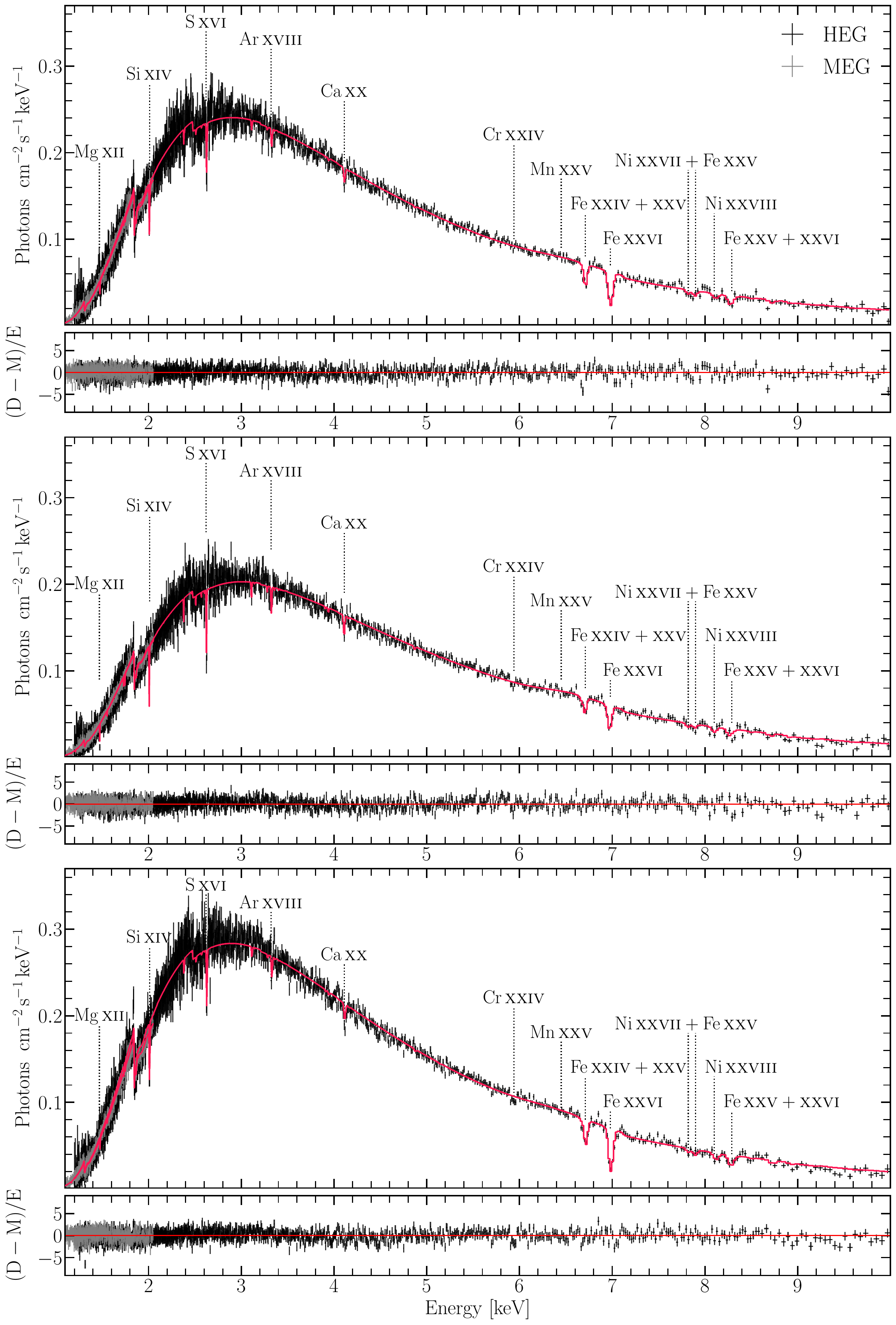}
\caption{HETG spectra of observations 20291, 20292, and 20293 (Epochs 1, 2, 4, from top to bottom). The best-fit model is shown in magenta. HEG (black data points) and MEG (gray data points) spectra were fitted simultaneously. Residuals from the best fit are displayed in the lower panel as $\rm (data-model)/error$. The HETG spectrum for Epoch 5 is shown in Figure \ref{fig:hetg}.}
\label{fig:all_hetg}
\end{figure*}

In our analysis, we fitted the X-ray spectra from the four \chandra/HETG observations corresponding to Epochs 1, 2, 4, and 5 separately. The continuum X-ray emission was modeled using two blackbody components (\texttt{bb} and \texttt{mbb} in \spex), while the accretion disk wind was described with a two-zone photoionization model (\pion). A Gaussian component was added to model the Fe K$\alpha$ emission line at 6.4-6.6 keV, and Galactic absorption was accounted for using neutral gas and dust components (\texttt{hot} and \texttt{amol}, respectively). The best-fit parameters are listed in Table~\ref{tab:results}. The \chandra/HETG spectra, along with the best-fit model (magenta line) and residuals, are shown in Figures~\ref{fig:hetg} and~\ref{fig:all_hetg}. These observations confirm the presence of accretion disk winds when the source is on the normal and flaring branches.

\begin{figure*}[t]
\includegraphics[width=\textwidth]{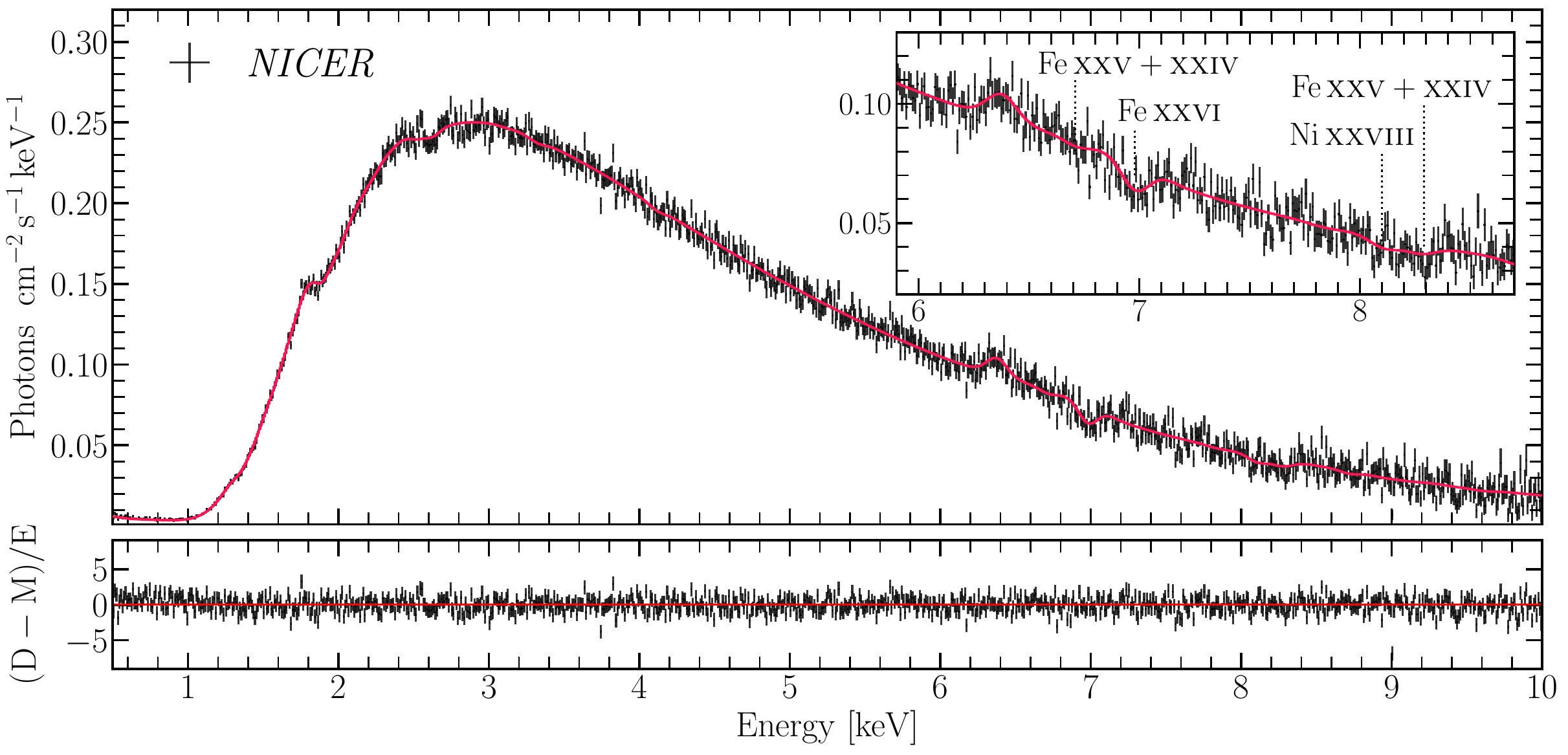}
\caption{\nicer spectrum of Epoch 3, with best-fit model in magenta. The residuals from the best fit are shown in the lower panel as $\rm (data-model)/error$. The inset highlights the Fe K band, where resonance absorption lines of the wind are detected.}
\label{fig:nicer}
\end{figure*}

During Epoch 3, the source was located on the horizontal branch. For this epoch, only a \nicer observation is available, as no \chandra data were obtained simultaneously. To investigate whether the accretion disk wind persists during the horizontal branch, we applied the same spectral model to the shorter \nicer observation from Epoch 3. The resulting spectrum and best-fit model are shown in Figure~\ref{fig:nicer}. Due to \nicer's lower energy resolution, only a single ionized absorption component could be resolved, with a column density of $\NH = (3.4 \pm 0.1) \times 10^{24}\:\rm cm^{-2}$ and ionization parameter $\log \xi = 4.8\pm 0.1$. The outflow and turbulent velocities were fixed at 500 km/s and 100 km/s, respectively. The temperatures of the \texttt{bb} and \texttt{mbb} components were measured at $1.55 \pm 0.03$ keV and $1.12 \pm 0.06$ keV, lower than the one of the other epochs. Column densities for the neutral gas, $\NH = (2.67\pm 0.04) \times 10^{22}\:\rm cm^{-2}$, and cosmic dust, $N_{\rm dust} = (1.40 \pm 0.07) \times 10^{18}\:\rm cm^{-2}$ were also estimated. Modeling of the Fe K$\alpha$ line required two Gaussian components: a broad emission line at $E_0 = 6.7 \pm 0.1$ keV with $\rm FWHM = 0.8 \pm 0.2$ keV, and a narrow line at $E_0 = 6.4 \pm 0.1$ keV with $\rm FWHM < 0.15$ keV. While the broad component is consistent with features observed in the HETG spectra, none of the HETG observations required the inclusion of a narrow component. A detailed analysis of the Fe K$\alpha$ line is beyond the scope of this study; a follow-up work will investigate its properties and evolution in relation to the mass accretion state, disk wind, and jet activity.

\section{\vla light curve and spectral indexes} \label{sec:appendix_vla}
\begin{figure*}[t]
\includegraphics[width=\textwidth]{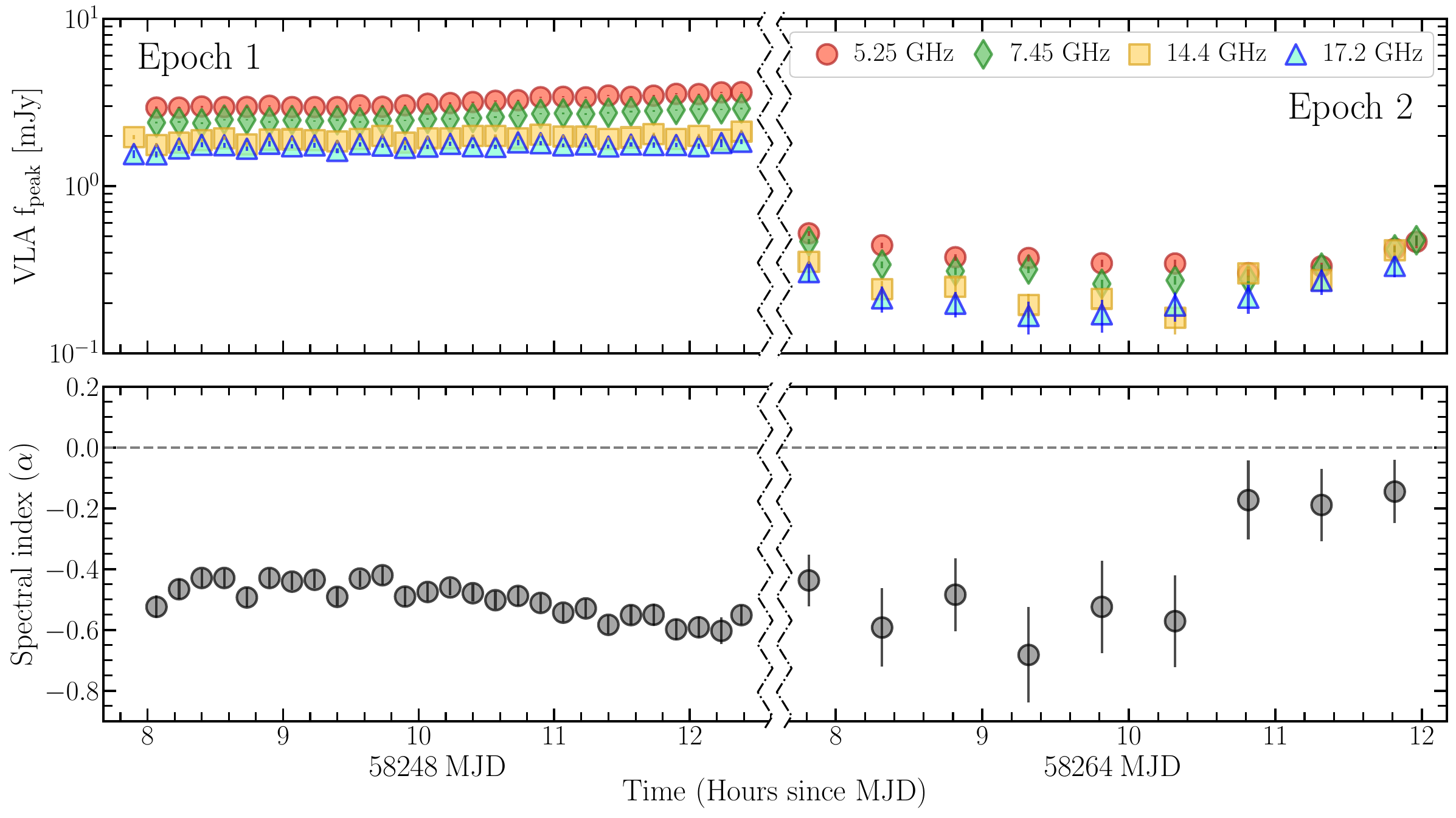}\\
\includegraphics[width=\textwidth]{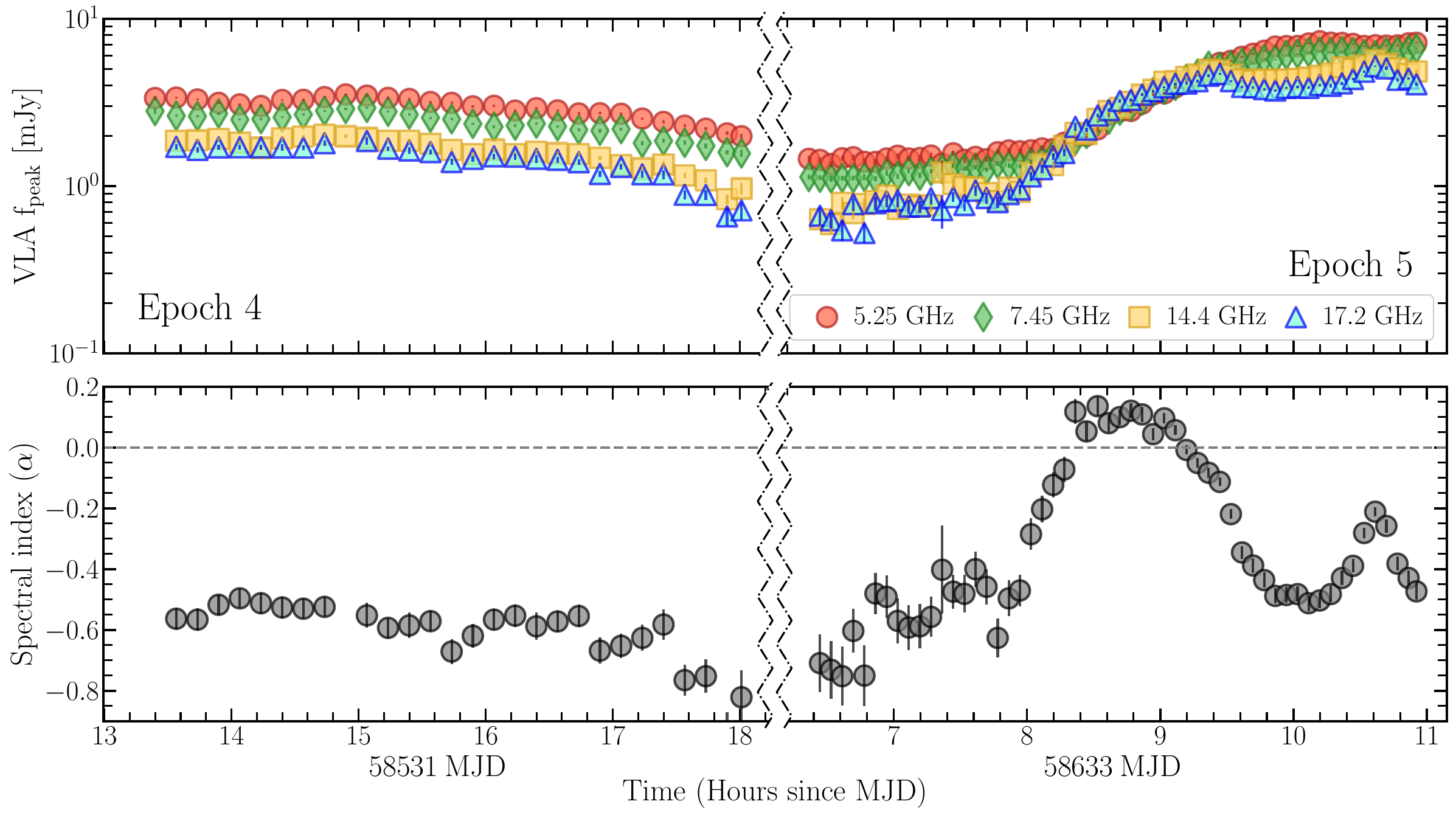}\\
\caption{5.25, 7.45, 14.4, and 17.2 GHz \vla light curves across different epochs. Below these, the corresponding spectral index, calculated across the four frequency bands, is displayed.}
\label{fig:vla}
\end{figure*}

In Figure \ref{fig:vla}, we displayed the full \vla light curve set, taken at four different frequencies: 5.25, 7.45, 14.4, and 17.2 GHz. The time evolution of the spectral indices is shown in the bottom panels.

\bibliography{biblio.bib}

\end{document}